\documentclass{aa}  

\usepackage{graphicx}
\usepackage{txfonts}
\newcommand{\ltsima}{$\; \buildrel < \over \sim \;$}
\newcommand{\simlt}{\lower.5ex\hbox{\ltsima}} 
\newcommand{\gtsima}{$\; \buildrel > \over \sim \;$}
\newcommand{\simgt}{\lower.5ex\hbox{\gtsima}} 

\newcommand{\feka}{\mbox{Fe  K$\alpha$}}
\newcommand{\fekb}{\mbox{Fe  K$\beta$}}

\newcommand{\xmm}{{\emph{XMM-Newton}}}

\newcommand{\lum}{erg~s$^{-1}$}
\newcommand{\flux}{{erg~cm$^{-2}$~s$^{-1}$}}
\newcommand{\nh}{cm$^{-2}$}
\newcommand{\nhsym}{N_{\mbox{\scriptsize H}}}

\newcommand{\sorg}{NGC~7582}
\newcommand{\chandra}{{\emph{Chandra}}}

\newcommand{\errUD}[2]{\ensuremath{^{+#1}_{-#2}}}
\newcommand{\errDU}[2]{\ensuremath{_{-#1}^{+#2}}}

\newcommand{\mgxi}{Mg\,\textsc{xi}}
\newcommand{\mgxii}{Mg\,\textsc{xii}}
\newcommand{\neix}{Ne\,\textsc{ix}}
\newcommand{\nex}{Ne\,\textsc{x}}
\newcommand{\Sixiii}{Si\,\textsc{xiii}}
\newcommand{\Sixiv}{Si\,\textsc{xiv}}
\newcommand{\Sxv}{S\,\textsc{xv}}
\newcommand{\Sxvi}{S\,\textsc{xvi}}
\newcommand{\Arxvii}{Ar\,\textsc{xvii}}
\newcommand{\Arxviii}{Ar\,\textsc{xviii}}

\newcommand{\fexxv}{Fe\,\textsc{xxv}}
\newcommand{\fexxvi}{Fe\,\textsc{xxvi}}
\newcommand{\suzaku}{{\emph{Suzaku}}}
\newcommand{\sax}{{\emph{BeppoSAX}}}

\newcommand{\logxi}{erg cm s$^{-1}$}
\newcommand{\nustar}{{NuSTAR}}

\begin{document}

\title{ A high  spectral resolution map of the  nuclear emitting  regions of NGC 7582 }

\author{V. Braito\inst{1,2}, J.~N. Reeves\inst{2,3},  S. Bianchi\inst{4}, E. Nardini\inst{3,5}, E. Piconcelli\inst{6}
}
\institute{
INAF - Osservatorio Astronomico di Brera, Via Bianchi 46 I-23807 Merate (LC), Italy; \email{valentina.braito@brera.inaf.it }
\and
Centre for Space Science and Technology, University of Maryland, Baltimore County, Baltimore, MD 21250, USA\\
\and Astrophysics Group, School of Physical and Geographical Sciences, Keele University, Keele, Staffordshire ST5 5BG, UK\\
\and Dipartimento di Matematica e Fisica, Universit\`a degli Studi Roma Tre, Via della Vasca Navale 84, I-00146 Roma, Italy \\
\and INAF -  - Osservatorio Astrofisico di Arcetri, Largo Enrico Fermi 5, I-50125 Firenze, Italy\\
\and INAF - Osservatorio Astronomico di Roma, Via Frascati 33, I-00040 Monteporzio Catone, Italy
} 

   \date{Received }
 \abstract
{We  present the   results of the spatial and spectral analysis of the  high resolution data of \sorg\, that were collected with a deep    ($\sim 200$ ksec) \chandra\ HETG  observation.  \sorg\ is a well studied ``changing look" AGN where multiple and possibly clumpy absorbers are present. During this long \chandra\ observation \sorg\ was in a highly obscured state. Therefore, we considered also a  short ($\sim 24$\, ks) \suzaku\ observation performed in September  2007, which caught \sorg\ in  a Compton thick state. This allows us to determine the underlying continuum model and  the amount of absorption [$\nhsym =(1.2\pm 0.2) \times 10^{24}$\,\nh].     A wealth of emission lines  (from Mg, Si, S and Fe) are detected in the \chandra\ spectrum, which   allow us to map the structure of the circum-nuclear emitters.  The high resolution spectrum reveals that the soft X-ray emission originates in a hybrid gas, which is ionized in part by the strong circum-nuclear star forming activity and in part by the central AGN. 
The new high resolution images confirm that the emitting region is highly inhomogeneous and extends up to a few hundred pc from the nuclear source.   The X-ray images are more extended in the lower energy lines (Ne and Mg) than in the higher energy lines (Si, Fe), where the former are dominated by the collisionally ionised gas from the starburst and the latter by the photoionized AGN emission. This is supported by the analysis of the He-like triplets in the grating spectra, where \mgxi\ displays strong resonance line emission (starburst) and where \Sixiii\ shows strong forbidden line emission (AGN). 
We deduce that  a low density ($n_e\sim 0.3-1 $ cm$^{-3}$)  photoionized gas is  responsible for the strong forbidden components and although it is more compact than the Ne and Mg emission, it is likely to originate from extended AGN Narrow Line Region gas at distances of 200-300 pc from the black hole.
We also detected an absorption feature at $\sim 6.7$ keV consistent 
with  the  rest frame energy of  the $1s-2p$ resonance absorption line from \fexxv \ ($E_{\mathrm{lab}}=6.7$ keV),
 which  traces the presence of a sub-parsec scale ionized circumnuclear absorber.   The emerging picture  is in agreement with   our new view of the circumnuclear  gas  in AGN, where    the medium is clumpy and stratified in both density and ionization. These absorbers and emitters are   located on different scales: from the   sub-pc   Broad Line Region gas out to the kpc scale of the galactic   absorber.}


\titlerunning{The HETG view of NGC~7582}
   \authorrunning{V. Braito et al.}
\maketitle

\section{Introduction} 
Systematic X-ray studies with \chandra\ and \xmm\     have established  that  our  classical view of the central regions of   Active Galactic Nuclei (AGN), where the absorbing medium is a homogeneous  pc scale absorber, covering a wide solid angle  is too simplistic (see \citealt{Bianchi2012,Turner2009} and references therein).   While we still believe that the central engine is fundamentally the same in every AGN (\citealt{Antonucci}), we have now mounting  evidence for a   more dynamic and complex circumnuclear gas.  Among the first  AGN that unveiled that complexity are the ``changing look'' AGN, which are extremely variable sources that transition from Compton-thin  (i.e. column density or $\nhsym <10^{24}$\,\nh), or even unobscured, to Compton thick ($\nhsym> 10^{24}$\,\nh) states  (\citealt{Matt2003,Guainazzi2002}). 
While for some of them a dramatic dimming  in the primary emission  can  explain the appearance as highly obscured, as the reprocessor can echo a past and more luminous emission (e.g NGC\,4051, \citealt{Guainazzi98}; NGC\, 2992, \citealt{Gilli2000}),  there is also a clear evidence that some of these extreme variations are due to  a highly inhomogeneous absorber. Variations of the covering factor or column density ($\nhsym$) of the X-ray absorbing gas,  with  the $\nhsym$ changing from few percent  up to an order of magnitude,    is a common property of obscured AGN  (\citealt{Risaliti2002,Markowitz2014,Torricelli-Ciamponi2014})  and argues for a high degree of clumpiness of the absorbing medium.  Although the majority of the absorption changes occurs on relatively long time-scales (i.e. months to years), we have witnessed $\nhsym$ variations on time-scales as short as few days or even  few hours, the prototype example of the class of ``changing-look" AGN being NGC 1365 (\citealt{Risaliti05,Risaliti07,Risaliti09, Maiolino2010}).   As some of these fast $\nhsym$ changes  can be regarded as eclipsing events, where our line of sight intercepts a higher density   cloud, their time-scales  can be directly linked to the size and distance of the absorbing clouds from the central     super-massive black hole (SMBH).   
These rapid  $N_{\rm H}$ variations  suggest   that    a significant fraction of this clumpy absorber must be located very close to the nuclear X-ray source and, more specifically,     within  the Broad Line Region (BLR; see \citealt{Risaliti05}).  
 
Intrinsic variability of the primary emission still plays a role as it is expected that the accretion  on to the   SMBH  to be   variable and characterized by flaring and dimming episodes, as shown by the  highly variable X-ray emission of genuine unobscured AGN (\citealt{McHardy1985,Green1993}).  The emerging picture of    AGN   is thus quite complex and the observed spectral variability cannot be solely ascribed to either intrinsic variability or the clumpiness of one single absorber.    Further insights on these multi phase and clumpy absorbers has come from  deep  observations of nearby   AGN with the  \xmm\ and \chandra\ gratings. These high spectral resolution  observations  revealed  that  the  X-ray emission  of AGN can be extremely rich in emission and absorption lines,  tracing  multiple emitting and absorbing   regions     all of them contributing to shape the observed  spectra (\citealt{Turner2009}).  In this framework ``Changing look"  AGN are an extremely important class  since they can place severe constraints on the location and structure of the X-ray absorbers. They are not  the exception to the AGN population, but most likely  they  are the sources  for which we have an optimal line of sight   that intercepts
all the innermost and clumpy   absorbers.\\

NGC\,7582  ($z=0.005254$) is an X-ray bright  ($F_\mathrm{2-10\, keV} \sim 3-8\times 10^{-12}$ \flux; \citealt{Bianchi09,Piconcelli2007}; hereafter  B09 and P07, respectively)  Seyfert 2 galaxy,  which has been observed  in the X-ray band several times: from the early days of X-ray astronomy  with \emph{Einstein} and \emph{Ginga} to the most recent X-ray observatories \xmm, \chandra\ and \suzaku. \sorg\ is classed as one of the prototype ``changing look" AGN as it has shown  remarkable  variability on different time-scales both in the optical  (\citealt{Aretxaga1999,Turner2000}) and in the X-ray band. Evidence for a variable  and complex X-ray  absorber  has been reported since the low spectral resolution observations performed with \emph{ASCA} and \sax\ (\citealt{Xue1998,Turner2000}). Furthermore, the first  broad band X-ray observation  of \sorg\ performed  by \sax\ (\citealt{Turner2000})  unveiled the presence of at least two X-ray absorbers, with one of them  being Compton thick and covering $\sim 60$\%  of the nuclear source.  A second and  Compton-thin absorber of $\nhsym \sim few \times10^{22}-10^{23}$\,\nh, part of which can be associated with a large scale  galactic absorber,   is always present and fully covers the nuclear source.
Subsequent  observations   confirmed  the extreme nature of NGC\,7582, where both flux and $\nhsym$ variations occur on different time-scales, from less than  a month to years.  
  On longer time-scales, such as the one probed with the two \xmm\ observations performed in 2001 and 2005 (P07),  
  the   thicker absorber increased in column density by    a $\Delta \nhsym \sim 7\times 10^{23}$\,\nh\, while the  thinner absorber remained constant ($\nhsym\sim 4\times 10^{22}$\,\nh). On  timescales as short as a day, as the one probed by two   observations performed with \xmm\  and \suzaku\  in April  2007 (B09),     the   thicker  absorber varies  by a typical    $\Delta N_\mathrm{H} \sim   10^{23}$\nh.  From the elapsed time  between those  two observations of about 20 hr,  B09 inferred a distance for the variable and higher $\nhsym$ absorber of about $few \times 10^{15}$ cm,  consistent with the location of the BLR. \\
   
    Given the rather constant Compton reflection component and   the   Fe K$\alpha$ emission line  detected in all the observations,  a larger scale Compton-thick reprocessor, that can be associated with the putative obscuring torus of the Unified Model of AGN, is also  present (B09).   Our line of sight  is not totally blocked by this material as we are able to view a  second    and more compact  absorber that   is responsible for the fast $\nhsym$ variations. The presence of  both  these absorbers  has been  recently confirmed by  two  \nustar\ observations that were performed  two weeks apart in 2012  (\citealt{Rivers2015}).     Finally,  a third  and  Compton thin   ($\nhsym \sim 10^{22}$\,\nh) absorber is  located at larger distance and could be identified with the   dust lane visible in  both  the \emph{Hubble   Space Telescope} (\emph{HST})   and Chandra X-ray images (\citealt{Bianchi07}, hereafter B07).  Interestingly, since the early  observations with \sax,    \sorg\ has  clearly undergone   a drop in the primary X-ray emission. Indeed, even in the Compton thin state the observed  hard (2--10 keV) flux  is  always of the order of $3-8\times 10^{-12}$ \flux\  and  it has never reached again the bright  state   observed with \sax\  ($F_\mathrm{2-10\, keV} \sim 2\times 10^{-11}$ \flux; P07, \citealt{Rivers2015}). \\

 Here we present a detailed analysis of a deep  ($\sim 200$ ksec) \chandra\ observation of \sorg, which was performed in April 2014 with the high energy transmission gratings    (HETG; \citealt{Markert94})  at the focal plane.      Prior to our   HETG observation,  \chandra\ observed \sorg\  with the 	Advanced CCD Imaging Spectrometer (ACIS; \citealt{Garmire2003})
 twice   in 2000 (for a total of $\sim 20$\,ksec),     which showed that the soft X-ray emission extends over $\sim 20 ''$  and  has a highly complex morphology. By comparing the \chandra\ and \emph{HST} images B07 concluded that the inhomogeneity seen in the X-ray images  can be in part explained with  the presence of the dust lane  of  the host galaxy and  in part  traces  variations of the physical properties of the emitting regions, which differ in ionization and/or density.   
    Furthermore, as seen in other Seyfert 2s   (\citealt{Guainazzi07}), the soft X-ray emission   is dominated by emission lines  from highly ionized elements (from He-like and H-like   O, Ne, Mg, Si and S),   with a strong contribution to   the emission from a photoionized gas (P07).  However, since  the \xmm-RGS  detector has a limited band pass  (0.35--2.5 keV) and  for \sorg\  the spectrum had insufficient signal to noise above  $\sim$1 keV,    most of the   emission lines were detected only with the \xmm\ and \suzaku\ CCD detectors.  Therefore the HETG observation presented here provides the first  high spectral resolution data of  this ``changing look"   over the  $1-10$ keV energy range.  During this observation \sorg\ was  in a highly obscured state, which allowed us to map the emitting regions up to the Fe-K band.      The paper is structured as follows: in \S 2  we describe the \chandra\   and \suzaku\ observations and data reduction; in \S 3 we present the  modeling of the \chandra\ spectrum.   In \S 4 we   compare the results of the spectral analysis with the  high spatial  resolution images extracted for the element of interest (i.e. Ne, Mg and Si) and derive the main   properties  of the multiphase emitting gas.  A summary of the structure of the circum-nuclear gas in \sorg\ is  presented in \S 4 and \S 5. \\

\section{Observations and data reduction}

\subsection {\chandra}
\chandra\  observed \sorg\   twice in 2014, on April 21 and April 22, for a total of $\sim 200$ ks (OBSIDs: 16079 and 16080 with  173 ks and 20 ks exposure, respectively; see Table~\ref{tab:logobs}). 
The observations were made  with the HETG instruments   at the focal plane of the Advanced CCD Imaging Spectrometer (ACIS-S; \citealt{Garmire2003}). The HETG consists of two gratings assemblies, the High-Energy Grating (HEG; 0.7--10.0 keV) and the Medium-Energy Grating (MEG; 0.4--8 keV). The HETG data were  reprocessed with the Chandra Interactive Analysis of Observations software package  (CIAO version 4.8\footnote{http://cxc.harvard.edu/ciao}; \citealt{Ciaoref}) and   CALDB version 4.7.2. 
We extracted the spectra   for each arm ($-1$ and $+1$) for the first order data of each of the HETG grating (HEG and MEG) for each observation.  To extend the HEG data above 8 keV we used an extraction strip narrower (with  a HETG  width of 20 instead of 35) than the default.  We then generated for each of the arms  ($-1$ and
$+1$) the spectral redistribution matrices ({\it rmf} files)  with  the  CIAO tool {\it mkgrmf} and the telescope effective area files with the  CIAO script {\it fullgarf}, which drives the  CIAO tool {\it mkgarf}. 
We then combined the    $-1$ and $+1$ orders (using response files combined with appropriate weighting);  we did not subtract the background  as it is negligible. 
  Although most of the  exposure  was performed during  the second   observation (OBSID 16079), we  checked  that  the spectra extracted from the two OBSIDs are consistent and combined  them to create  higher quality summed first-order HEG and MEG spectra for fitting; again we created the   response files combining them with the appropriate weighting.   Prior to any spectral fitting it is clear that \sorg\ was in a Compton-thick state for the whole duration of the \chandra\ observation, as  there is  no evidence of the primary emission emerging in the collected spectra up to 5 keV (see Fig.~\ref{fig:eeuf_suzaku_chandra}).   For the   main spectral fits  we binned the HEG to 1024 channels and the MEG to  2048   channels, corresponding to  $\Delta \lambda \sim 0.02\AA$. This  binning corresponds to twice the full width half maximum (FWHM)  resolution of the HEG and  the FWHM of the  MEG detectors.  A finer binning of  $\Delta \lambda \sim 0.01\AA$ will be used later for some of the line profiles.  For most of  the spectral fits we considered the 1.2--7 keV and 1.9--9.5 keV energy range for the MEG and HEG, respectively;  in these energy  bands  we collected $\sim 1140$ and $\sim 1105 $ counts in the MEG and HEG spectra. 
Throughout the paper  we adopted  H$_0=71$ km s$^{-1}$ Mpc$^{-1}$, $\Omega_{\Lambda}$=0.73, and $\Omega_m$=0.27 \citep{Spergel2003}. 
All the  spectral fits are performed using \textsc{XSPEC} (\citealt{xspecref}) with the C-statistic; the  fit parameters are quoted in the rest frame of \sorg\ and errors are at the 90\% confidence level   for one interesting parameter, unless otherwise stated.  As we are using the   \textsc{XSPEC} software package, all fit parameters are given in energy, where we adopted the conversion factor   between energy and wavelength of   1 keV being  $12.3984$\, \AA\  in wavelength.  In all the fits we included   Galactic absorption   with a column density $\nhsym=1.9 \times 10^{20}$   (\citealt{Dickey}). \\

\begin{table}
\caption{Log of the \chandra\ and \suzaku\ observations of \sorg\ considered for the spectral analysis.}
\begin{tabular}{lcccc}
\hline\hline
 Obs.     &  Det. & OBSID     & DATE &  $T_\mathrm{EXP}$ \\
  &   & & &   {(ks)} \\
\hline
 \chandra\ & HETG & 16080 & 2014 Apr 21& 19.7 \\
\chandra\  & HETG & 16079 & 2014 Apr 22 & 173.4 \\
\suzaku${^\mathrm{a}}$  & XIS & 702052040 & 2007 Nov 16 & 23.9\\
 \suzaku${^\mathrm{a}}$ & PIN & 702052040 & 2007 Nov 16 & 23.9\\
\hline
\end{tabular}
\tablefoot{
\tablefoottext{a}{Net exposure time after standard cleaning.} 
}
 \label{tab:logobs}
\end{table}

\subsection{\suzaku}
Four  \suzaku\ \citep{Mitsuda07} observations of \sorg\ were performed in 2007     with an averaged exposure time of 25-30   ksec. As  \sorg\ was Compton  thin in the first two observations and thick in the last two (B09),  we  considered only two  observations representative of each state; in particular we  considered for the thick state the fourth  observation, which has an observed 2--10 keV flux similar to the flux measured with \chandra\ ($F_{2-10\,\mathrm{keV}}\sim 3 \times 10^{-12} $\,\flux, see Table~\ref{tab:cont}). \\  
For the \suzaku\ data reduction we followed the standard prescription  for the filtering procedures and considered only the  data from the  two front illuminated (FI)  X-ray Imaging Spectrometer  (XIS 0 and XIS 3, \citealp{Koyama07}) and the hard X-ray  detector (HXD-PIN; \citealt{Takahashi}).   The XIS    source spectra  were extracted from a circular region of 2.9$'$ radius  centered on the source,  while the  background spectra  were extracted from two circular regions of 2.4$'$ radius  offset from the source and   the Fe 55 calibration sources, which are in two corners of the  CCDs.  
The XIS response   and ancillary response   files were produced,   using the latest calibration files, with the \textit{ftools} tasks \textit{xisrmfgen} and \textit{xissimarfgen}, respectively.   After checking for consistency, the spectra from the  two  FI-CCDs  were combined  in  a single source spectrum.    
The HXD-PIN data  were reduced following the   latest \suzaku\ data reduction guide (the ABC guide Version
4.0\footnote{http://heasarc.gsfc.nasa.gov/docs/suzaku/analysis/abc/}),   and using the rev2 data, which
include all 4 cluster units.   The HXD-PIN  background was extracted from the background event file (known as  the ``tuned'' background), which is provided by the HXD-PIN  team and accounts for  the instrumental  Non X-ray Background (NXB; \citealp{kokubun}).  The source and background spectra were extracted  using the same common good time interval and   the source spectrum  was corrected for the detector dead time.  We  then  simulated a spectrum for the cosmic X-ray background counts  adopting the form of  \citet{Boldt} and \citet{Gruber} and the response matrix for the diffuse emission; the resulting spectrum was then added  
to the  instrumental one.\\ 
  
  The   XIS-FI source spectra were   binned  to 1024 channels and then   to a  minimum  of   50 counts per bin,   while the HXD-PIN spectrum was   binned in order to have a signal-to-noise ratio of 5 in each energy bin. 
  The net count rates   for the Compton thick state are: $( 0.067\pm 0.001)$ counts s$^{-1}$   and  $( 0.052\pm 0.003)$ counts s$^{-1}$ for the XIS-FI (0.6--10 keV)  and HXD,  respectively.    Data were included from 0.6--10 keV for the XIS--FI,    we also excluded the data in the 1.6--1.9 keV energy range due to calibration uncertainties; for the HXD-PIN we considered only the 14--50 keV energy range.

 \begin{figure}
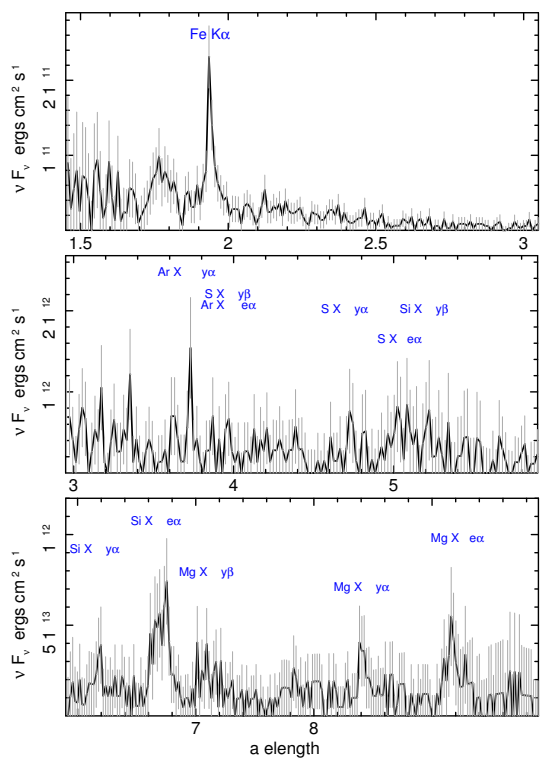

 \begin{center}
\includegraphics[angle=-90,width=0.4\textwidth]{new_eeuf_heg2048nolab.eps}
\includegraphics[angle=-90,width=0.4\textwidth]{eeuf_meg_3-6nolab.eps}
\includegraphics[angle=-90,width=0.4\textwidth]{eeuf_meg_6-10.eps}
 \caption[wav_spec]{  \chandra-HETG spectrum of \sorg, shown  versus the rest-frame wavelength (increasing in $\lambda$ from the top to the lower panel). The spectra are binned at the FWHM resolution for both the HEG ($\Delta \lambda=0.01$\, \AA) and the MEG ($\Delta \lambda=0.02$\, \AA). The spectra have been folded against a power-law model with $\Gamma =2$. The top panel shows the HEG spectrum in the $1.45-3\,$\AA\, range (corresponding to $\sim 4-8.5$ keV) covering the \feka\ emission line. The middle and lower panels are from the MEG grating.    The strongest emission features  (labeled in blue)  are due to Argon  (at $\sim3.7$\, \AA), Sulphur (at $\sim 5 $\, \AA\, range), Silicon, with the prominent He-like triplet at $\sim 6.6$\,\AA,  and   Magnesium (in the  $\sim 7-9$\, \AA\, range). The broadband spectrum  is line dominated  with an extremely weak continuum emission, which emerges  only below  $\sim 2.5$ \,\AA\ (i.e. at $E\simgt 5$\, keV). 
  \label{wav_spec}
}
\end{center}
\end{figure}
\section{Spectral Analysis}

\subsection{Continuum  modeling}
The first inspection of the  \chandra\  data shows a spectrum dominated  by emission lines with an extremely low continuum,
suggesting that  NGC~7582 was in  a  Compton thick state  for the  duration of the observation. In Fig.~\ref{wav_spec} we show the observed HETG spectrum in $\nu F_\nu$ flux units unfolded against a power-law model with a photon index $\Gamma=2$.  The spectrum is shown at the FWHM resolution of the  HETG  ($\Delta\lambda =0.01$\,\AA\  and $\Delta\lambda=0.02$\AA\ for the HEG and MEG, respectively) and covers the wavelength range from $1.45-10$\, \AA\ (corresponding to $1.25-8.5$ keV).  The spectrum is plotted in  wavelength to  show not only the  overall  spectral shape but also the several emission lines that are present.
 Before proceeding to parameterize the emission lines,  we  focused on the  shape of the underlying continuum. As at this point we are not interested on the finer details of the emission line profiles, we   considered  the HEG and MEG spectra  grouped with a minimum of 10 counts   per bin and   fitted in  the 0.7--7 keV  and  1.9--9.5 keV energy range for the MEG\footnote{We note that adopting the additional grouping of 10 counts allows us to consider the MEG data down to  0.7 keV.} and the HEG, respectively.   We considered a simple baseline continuum model  typical of a highly obscured AGN, which is  composed of an absorbed power-law component,  a scattered soft power-law  and a  Compton reflected continuum  component (modeled with \textit{PEXRAV}, \citealt{pexrav}), where we assumed an  inclination angle of  $45^{\circ}$,  a reflection fraction of $R=1$   and  allowed its normalization to vary with respect to the primary power law.    We assumed that the scattered  and primary power-law component have the same photon index ($\Gamma$). This  baseline continuum model provides a reasonable description of the overall spectral shape; the fit statistics is poor  ($C=463.2$ for 171 $d.o.f.$) and even at this coarse binning   several line-like residuals are clearly present.    Due to the limited photon statistics, as well as the lack of  coverage above 10 keV,  as we are not including the \suzaku\ data,  both the intrinsic photon index and  the amount of absorption are  unconstrained. However,     we can still derive a lower limit on the amount of absorption of  $N_\mathrm{H}> 1.2\times 10 ^{24}$\,cm$^{-2}$,   assuming a photon index $\Gamma=1.8$, in agreement with  a Compton-thick scenario. \\

\begin{figure}
\centering
\includegraphics[angle=-90,width=0.45\textwidth]{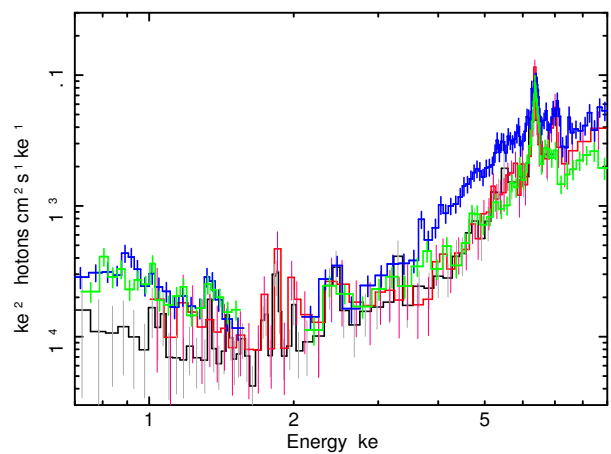}
 \caption[eeuf_suzaku_chandra]{ Comparison between the \chandra\ (HEG and MEG; red  and black data points, respectively) and two of the  \suzaku\ observations,  which caught this AGN in  the two different states  representative of the Compton thin ($\nhsym \sim 4\times 10^{23}$\, \nh; blue data points) and thick ($\nhsym \sim 1.2\times 10^{24}$\, \nh;  green data points) states.   Above  2 keV the \chandra\  spectra are similar in  observed flux and spectral shape to the typical Compton thick state observed by \suzaku.  In the soft X-ray band the lower flux is due to the smaller extraction region of the \chandra\ grating data. The spectra are unfolded against a power law with $\Gamma =2 $ and  rebinned for plotting purposes.  
 \label{fig:eeuf_suzaku_chandra}
 }
\end{figure}

 \begin{table}
\caption{Best fit parameters  and observed fluxes for the continuum derived with the joint fit of the \suzaku\ and \chandra\ data }
\begin{tabular}{llc}
\hline\hline
Parameter    &  Units &  \\
 \hline
  $\Gamma $& &  1.74\errDU{0.18}{0.18}\\
 $\nhsym$ & $10^{22}$\, \nh & 120\errDU{18}{21} \\
 A$_\mathrm{Scatt}^\mathrm{a}$  & $10^{-5}$\,ph cm$^{-2}$ s$^{-1}$ & 7.6\errUD{0.6}{0.6} \\
  A$_\mathrm{PL}^\mathrm{b}$ &  $10^{-3}$\,ph cm$^{-2}$ s$^{-1}$ & 3.1\errDU{1.2}{2.4} \\
 A$_\mathrm{PEX}^\mathrm{c}$ & $10^{-3}$\,ph cm$^{-2}$ s$^{-1}$ & 1.9\errDU{0.9}{1.2}  \\
 kT & keV & 0.63\errUD{0.05}{0.05}\\
 A$_\mathrm{Th_{Suzaku}}^\mathrm{d}$  &$10^{-4}$ cm$^{-5}$  & 1.20\errDU{0.20}{0.22}\\
  A$_\mathrm{Th_{Chandra}}^\mathrm{e}$  &  $10^{-4}$\,cm$^{-5}$ & 0.37\errDU{0.11}{0.12}\\
  F (0.5--2) keV$_\mathrm{Chandra}$ & $10^{-13}$ \flux & $1.9\pm0.1$ \\
  F (0.5--2) keV$_\mathrm{Suzaku}$ & $10^{-13}$ \flux & $4.3\pm0.2$\\
  F (2--10) keV$_\mathrm{Chandra}$ & $10^{-12}$ \flux & $3.3\errDU{0.7}{0.2}$ \\
  F (2--10) keV$_\mathrm{Suzaku}$ & $10^{-12}$ \flux & $3.1\errDU{0.5}{0.1}$ \\

\hline
\end{tabular}
\tablefoot{
\tablefoottext{a}{Normalization of the scattered power-law component.}\\
\tablefoottext{b}{Primary  power-law component normalization.}\\
\tablefoottext{c}{Compton reflection component   normalization.}\\
\tablefoottext{d}{Normalization of   the thermal component  is defined as $10^{-14}/[4\pi(D_{A}^2 \times (1 + z)^2]) \int n_e n_H dV$, where $D_A$ is the angular size distance to the source (cm), $n_e$ is the electron density (cm$^{-3}$), and $n_H$ is the hydrogen density (cm$^{-3}$) as defined in the {\sc mekal} model.}\\ 
\tablefoottext{e}{Normalization of   the thermal component   for the \chandra\ spectra obtained  after fixing the photon index of the scattered component. }
}
 \label{tab:cont}
\end{table}

A soft excess, possibly associated to  a starburst emission  is present below 1 keV;  however, while   the addition of  a  thermal emission component (\textsc{mekal}; \citealt{Mewe85}) improves the fit ($C=407$ for 169 d.o.f.) several residuals are present as most of the line-like  features   that are visible in Fig.~\ref{wav_spec} are not accounted for by this model.   The temperature of this component is  $kT =0.78^{+0.06}_{-0.16}$\, keV and mainly improves the fit  below 1 keV,   but cannot account for  both the strong   \mgxi\ and \Sixiii\  emission lines. We tested for the presence of an additional   and higher temperature thermal  emission component, but also in this case while the fit statistically improves ($\Delta C=25$ for 2 d.o.f.) it still leaves strong residuals above $1.2$\, keV.  As we will show later   there is evidence  in the zeroth order images for some extended emission, which    extends on scales of few hundreds pc from the nuclear source. This   could  be,   in principle, associated with  the Narrow Emission Line Region (NLR) gas or a   hot gas   connected with a strong circum-nuclear star formation activity.    Most of the emission  lines    with the exception of \mgxi\ show  also a strong forbidden component, which is indicative of a photoionised gas  or at least an  hybrid plasma.   Therefore the simple addition of a thermal emission model cannot describe adequately the soft X-ray spectrum. This is in agreement with the analysis of the XMM-RGS data presented by P07, where, despite the larger aperture of the RGS data   with respect to the \chandra\ gratings, it was already found that   the soft X-ray emission in NGC~7582 has a strong contribution from   a photoionised gas. \\

 In order to better  parametrize the continuum,  we then repeated the fit including both the \chandra\ and the
   \suzaku\ spectra collected   during the fourth   of   the \suzaku\ observations (both the XIS-FI and the HXD-PIN data), which caught \sorg\ in a Compton thick state and with a similar 2--10 keV flux.    As we are not interested in the detailed modelling of the absorber, we fitted  the same simple  model  for a Compton thick absorber/reflector and we included  several  Gaussian lines  to account for the soft X-ray emission lines as well as the \feka\ emission line.   Since the  \suzaku's  large PSF  (point spread function)  includes all the galaxy,  we added a thermal emission component for the \suzaku\  data allowing for   different normalizations between the \chandra\ and   \suzaku\  spectra. Indeed  a clear  excess in the soft X-ray emission is  apparent in the Suzaku XIS below 1 keV  with respect to the \chandra\  data  (Fig.~\ref{fig:eeuf_suzaku_chandra}). However, as  the \chandra\ spectra have a low   S/N  below 1 keV, the normalisation   of  the  \chandra's thermal component    is highly degenerate with the photon index and normalisation of the scattered power-law component, so   we    initially  fixed it  to zero.  This model can be expressed with the mathematical form:
   \begin{equation}
   \begin{split}
    F(E)= \mathrm{abs_{\mathrm{Gal}}\times (pow1 + zphabs\times pow2  +} \textsc{pexrav} +\\
    \mathrm{ Gaus_{soft} +  Fe \,K\alpha   +\textsc{mekal})},
   \end{split}
    \end{equation}
    where as before we assumed the same $\Gamma$ for both the scattered  (pow1) and primary power-law component (pow2). Thanks to the joint fit with the \suzaku\ data we  have  now some basic constraints on the primary continuum:  the photon index is found to be $\Gamma=1.74\pm 0.18 $, the column density of the absorber is, as expected, in the Compton thick regime   with   $\nhsym=1.2\pm 0.2 \times 10^{24}$\,\nh,  the normalisations of the primary and reflected  power-law components are  $N_{\mathrm{PL}} =3.1\errUD{2.4}{1.2} \times 10^{-3}$ photons cm$^{-2}$ s$^{-1}$   and   $N_{\mathrm {PEXRAV}} = 1.9\errUD{1.2}{0.9} \times 10^{-3}$ photons cm$^{-2}$ s$^{-1}$ (see Table~\ref{tab:cont}), which correspond to a reasonable  reflection fraction of  about $R\sim 0.6$.   These parameters are   in agreement with the  best-fit model of the Compton thick state as found by B07. Although statistically the fit is still not  good ($C/d.o.f=402.1/238$),  we note that most of the remaining residuals   are  mainly due to additional soft X-ray emission lines in the \chandra\ spectra. The observed   \chandra\ soft  (0.5--2 keV) and hard (2--10 keV)  X-ray  fluxes are  $F_{(0.5-2)\mathrm{keV}}\sim 1.9 \times 10^{-13} $\,\flux\ and $F_{(2-10)\mathrm{keV}}\sim 3.3 \times 10^{-12} $\,\flux\, respectively (see Table\, 2). We note that the  hard X-ray  flux  measured with this observation  is similar to all the recent observations  that caught NGC\,7582 in a thick state  (B09; P07; \citealt{Rivers2015}). \\

    \sorg\ was also observed with \nustar\ in 2012 on August 31 for $\sim  16 $\, ksec    and September 14th for $\sim 14$\, ksec (\citealt{Rivers2015}).  We inspected both these observations and found that while in the first observation  \sorg\ was caught in the standard Compton thin state ($\nhsym \sim 2\times 10^{23}$\,\nh), in the second   \sorg\  was in a more obscured  state ($\nhsym \sim 6\times 10^{23}$\,\nh). We attempted a simultaneous fit with the \chandra\ and \nustar\ spectra as done with the \suzaku\ data;  however, while  in the 2--10 keV energy range  the \suzaku\ spectra perfectly overlay on the \chandra\ data (as seen in Fig.~1), the \nustar\ spectra show a  slightly different spectral shape.  This could be  explained with  either a  different $\nhsym$ and/or an apparent harder photon index.    Therefore we proceeded to adopt the best-fit continuum described by the joint analysis of the \chandra\ and \suzaku\ spectra in the Compton thick state (see Table~\ref{tab:cont}).\\

  \subsection{Emission line component}
Having reached a good representation of the underlying continuum  we then proceeded to build the \chandra\ emission line component model. We  therefore fixed the photon index and $\nhsym$ to the  best fit value found with the joint fit with the \suzaku\ spectra  (see Table~\ref{tab:cont}) and allowed  the normalisations of the power-law components and reflection component to adjust  upon adding the Gaussian emission lines. We considered at first the MEG and HEG  data binned at $\Delta \lambda =20$\,m\AA\  (2048 and 1024 channels for the MEG and HEG respectively) and the $1.2-7$ keV  and  $1.9-9.5$ keV  energy range for the MEG and the HEG spectra, respectively.   
 We then added several Gaussian  emission lines;  each individual line was considered to be statistically significant if the addition yielded an improvement of the fit statistic of $C > 9.2$ (corresponding to 99\% significance for 2 interesting parameters), with the exception of possible weaker emission lines that are part of a triplet (see below).  We then allowed their widths to be free; if the width was found to be  consistent with being unresolved at the 90\%  confidence level,  we then fixed it  to $\sigma=1$\,eV.   Below 6 keV  we detected 10  statistically significant Gaussian emission lines, whose main parameters together with the most likely identification, the improvement of the fit and laboratory energy  are listed in Table~\ref{tab:line1}. The parameters  of four  less significant lines are  also listed as they correspond to the expected  transitions    from \mgxi, \mgxii\ and \Sxvi. After the inclusion of these  Gaussian emission lines,  any addition of a thermal emission component did not improve the fit. This is because  we considered the data only above 1.2 keV  and once we accounted for the emission lines no spectral curvature (i.e. from an additional weak thermal emission component) is left unmodeled.   This does not imply that we can rule out the presence of a thermal emission component, indeed  as  we will show later some of the emission  lines could be explained with a collisionally ionized plasma.  \\
 \begin{figure}
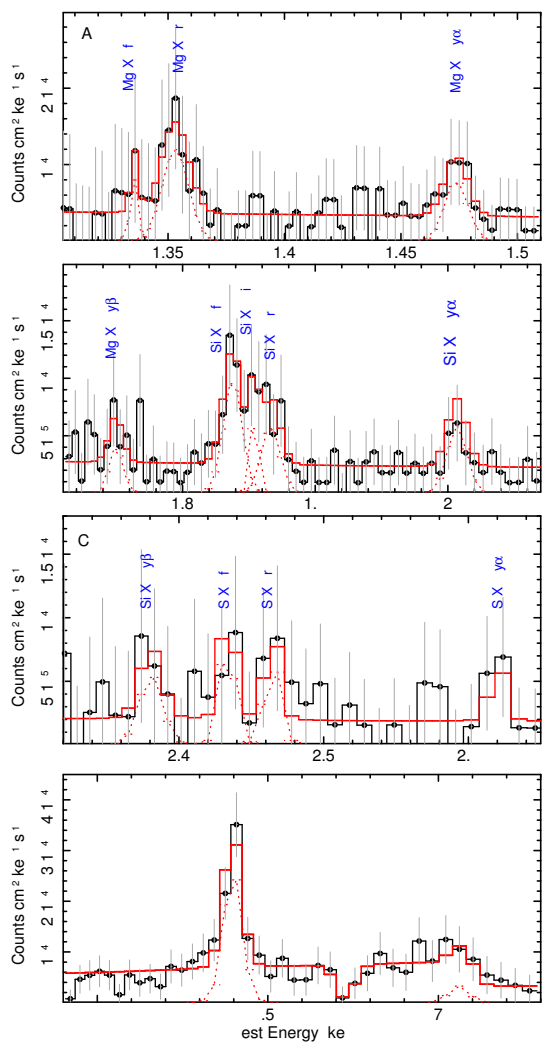

 \begin{center}
\includegraphics[angle=-90,width=0.4\textwidth]{new_MgXII_nolab.eps}
\includegraphics[angle=-90,width=0.4\textwidth]{new_SiB_nolab2.eps}
\includegraphics[angle=-90,width=0.4\textwidth]{Si_S_nolab.eps}
\includegraphics[angle=-90,width=0.4\textwidth]{newFe_2048.eps}
 \caption[mulitplot-2048]{Zoom into the MEG  (top 3 panels, labeled as A, B and C) and HEG spectra (lower panel)  showing the strongest emission line features.  The spectra are binned at the $FWHM$ resolution  for  both the MEG and the HEG. The  main Gaussian emission lines are labelled and shown with a red dashed line. Panel A shows the \mgxi,  and \mgxii\  Ly$\alpha$ region.  In panel B we show a zoom into the  region of the \mgxii\   Ly$\beta$, the  \Sixiii\ triplet  and the \Sixiv\ Ly$\alpha$, while  panel C shows  the \Sxv\ and \Sxvi\  range. In panel D we show the \feka\ region, beside the \feka\ line at $\sim 6.4$\, keV a possible weak absorption feature is present at $\sim 6.7$ keV.
 \label{fig:2048_plot}
}
\end{center}
\end{figure}

\begin{table*}
\centering
\caption{Emission lines  derived from the fit to the 2048 channel MEG and 1024 channel HEG spectra.   The statistic for the model with no lines is $C/d.o.f. = 1122.9/683$. In the last column we  list the improvement of the fit upon adding each of the  emission lines. }\begin{tabular}{ccclcc}
\hline\hline    
Rest Energy$^{\mathrm{a}}$     &Intensity & $\sigma$ & ID$^{\mathrm{b}}$      & Atomic Energy$^{\mathrm{c}}$& $\Delta C^{\mathrm{d}}/d.o.f$\\
(eV) & ($10^{-6}$ph cm$^{-2}$ s$^{-1}$)&(eV) & & (eV) & \\
\hline  
$1336\errDU{4}{3}$   & $ 0.36\errDU{0.32}{0.48}$ &$1^{\mathrm{f}} $& \mgxi\ (f)& 1331 & 3.9/2\\  
$1353\errDU{3}{3}$   &  $1.7 \errDU{0.7}{0.8}$ &  $5.4\errUD{2.6}{3.5}$  & \mgxi\ (r)& 1352 & 32.5/3\\
$1474\errDU{4}{3} $  & $1.0  \errDU{0.5}{0.6}$ & $4.7   \errDU{2.6}{8.4}$ &  \mgxii\ Ly$\alpha$ &1473&  18.6/3 \\
 $1750\errDU{5}{9}$  & $0.55 \errDU{0.35}{0.48}$ & $4.7^{\mathrm{t}} $  & \mgxii\ Ly$\beta$  &1745& 6.5/2 \\
 $1838 \errDU{3}{4}$ & $1.8 \errDU{0.6}{0.8} $ & $6.5\errDU{2.4}{4.5}$  &\Sixiii\ (f)  &1839 &49.1/3\\ 
$1866\errDU{8}{5}$ &  $1.1\errDU{0.5}{0.9}$ & $6.5^{\mathrm{t}} $  &  \Sixiii\ (r)   &1865& 33.3/2\\
$ 2007 \errDU{4}{4}$   &$1.0\errDU{0.4}{0.6}$ &$6.2\errDU{3.6}{6.8}$&  \Sixiv\ Ly$\alpha$ &2006& 23.5/3\\
$2381\errDU{11}{5}$ & $1.1\errDU{0.6}{0.8}$ & $6.2^{\mathrm{t}} $    &    \Sixiv\ Ly$\beta$ &2376& 9.4/2\\
$2434\errDU{3}{3}   $ &$1.1\errDU{0.6}{0.8}$   &$1^{\mathrm{f}} $ &  \Sxv\ (f) &2430&14.2/2\\
$2465\errDU{3}{6}$& $0.90\errDU{0.52}{0.71}$ &$1^{\mathrm{f}} $ &  \Sxv\ (r) &2461&9.9/2\\
$2620\errDU{5}{10} $ &$0.65\errDU{0.46}{0.65}$&  $  1^{\mathrm{f}} $ &  \Sxvi\ Ly$\alpha$ &2622&5.4/2\\
 $3104\errDU{10}{5} $   &$0.60\errDU{0.37}{0.51}$ &$  1^{\mathrm{f}} $  &  \Sxvi\ Ly$\beta$ &3106&7.1/2\\
  &                                          & & \Arxvii\ (f)& 3104 &\\
$3325\errDU{6}{5} $  & $0.9\errDU{0.46}{0.59}$  &$  1^{\mathrm{f}} $   & \Arxviii\ Ly$\alpha$& 3323 &17/2\\ 
 $6390\errDU{13}{11}$ & $ 17.3 \errDU{0.4}{0.4}$  &$23\errDU{14}{9}$ & Fe K$\alpha_1$&  6390  & 112.5/3 \\
 &                                          &  & Fe K$\alpha_2$&  6403 & \\
\hline
\end{tabular}
\tablefoot{
\tablefoottext{a}{Measured line energy in the  \sorg\ rest frame.} \\
\tablefoottext{b}{Possible identification.}\\
\tablefoottext{c}{Known atomic energy of the most likely identification of the  line in eV, from NIST (http://physics.nist.gov/PhysRefData/ASD/).}  \\
\tablefoottext{d}{Improvement in C-statistic upon adding the line.}\\
\tablefoottext{f}{Denotes the parameter is fixed.}\\
\tablefoottext{t}{Indicates that the $\sigma$ is tied to the width of another transition from the same ion species}\\
}
 \label{tab:line1}
\end{table*}
 In  Fig.~\ref{fig:2048_plot} we show a zoom into the  range of interest of the strongest emission lines, noting that no other clear residuals are left; furthermore   the plots show that the baseline continuum,  derived from the joint fit with the \suzaku\ spectra, is a good representation of the continuum in the HETG spectra. In particular, as can be seen in the lower panel (at the Fe-K region) no further residuals due to a reflection edge are present.
 We then  examined, when possible, each of the strongest emission lines complexes with  the spectra binned at Half Width at Half Maximum (HWHM corresponding to  4096 channels for the MEG) to investigate the actual broadening of some of the emission lines that appear to be resolved and  derive  a more general scenario for the emitting regions.  All the emission lines  appear to be detected at the  expected rest-frame energy; in order to quantify any possible in- or outflow velocity  of the emitting gas we considered  the 90\% errors on the energy centroids of  the strongest H-like emission lines (see Table~\ref{tab:line1} and  Table~4). We     found that although the energy centroids  may suggest outflow velocities of $\sim 200-800$ km s$^{-1}$, once we consider the large errors ($\pm 1000$ km s$^{-1}$) they are all consistent with  the systemic velocity of \sorg.

\subsubsection{The \mgxi\  and  \mgxii\ emission lines }  
 At  FWHM resolution binning, in the energy range of the \mgxi\ triplet we clearly  detected  only  the resonance component at $E=1353\pm 3$\, eV,  a weaker emission line may be present at the corresponding energy of the forbidden  component, but it is not statistically significant ($\Delta C <4$ for 2 d.o.f.).  The resonance component is  resolved with a  broadening of $\sigma=5.4\errUD{2.6}{3.5}$ eV corresponding to a $FWHM \sim 2800\pm 1600$ km s$^{-1}$;   forcing it to be unresolved  (i.e. 1 eV)  the fit worsens by  $\Delta C=9.3$.  The \mgxii\ Ly$\alpha$ line is also clearly detected  ($\Delta C=18.3$) with an $EW=30\errDU{14}{19}$ eV  and  is also    resolved with a $FWHM \sim 2300$ km s$^{-1}$ ($\Delta C= 6$  with respect to  $\sigma=1$eV).  
 
\begin{figure}
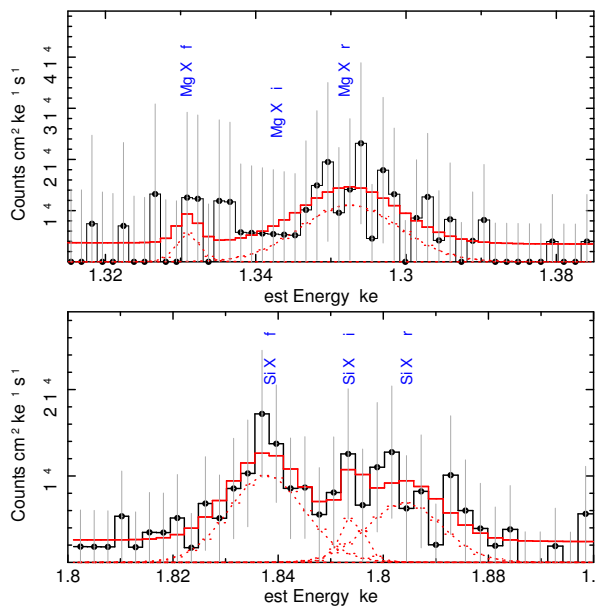

\includegraphics[angle=-90,width=0.44\textwidth]{MgXII_4096.eps}
\includegraphics[angle=-90,width=0.455\textwidth]{Si_XIII_4096.eps}
\caption[MgXII_Si_4096]{Zoom   into the MEG spectrum  at the energy of the \mgxi\   and \Sixiii\  triplets with the 4096 channels binning (i.e. at  $HWHM$ of the spectral resolution). The blue   labels mark the expected position of the forbidden, intercombination and  resonance transitions.  The top panel shows the \mgxi\  region.  While there is no evidence of emission at the energy of the forbidden   component, the finer binning  confirms both the strength and broadening  of the resonance  emission line.  The lower panel shows the  \Sixiii\  triplet, where  the finer binning  confirms both the strength and broadening  of the forbidden and resonance  emission lines. While it is not statistically required   ($\Delta C=2$) we included also a Gaussian emission line to account for a possible  intercombination component.
\label{fig:MgXI_Si_4096_plot}
}
\end{figure}

 We then considered the MEG spectrum binned at  the HWHM  (i.e. 4096 channels,  corresponding to  a spectral resolution of $\Delta \lambda \sim 0.01\AA $) and  refitted the \mgxi\ and \mgxii\ lines.  The energy centroids    and normalisations of the   \mgxi\ and \mgxii\ emission lines are similar to the values obtained with the coarse binning and also  at this finer binning the forbidden component is not statistically required and we can  only derive  an upper limit on its intensity of  $I<6.5 \times 10^{-7}$ ph cm$^{-2}$ s$^{-1}$. 
 Both the  strong resonance component  of  the \mgxi\ and the \mgxii\  Ly$\alpha$ emission lines are  resolved  (see Fig.\ref{fig:MgXI_Si_4096_plot}, upper panel  and Table~\ref{tab:line2}),  with a possible broadening of $\sigma\sim 6$\, eV,  which corresponds to    $ FWHM$  velocity widths  of   $3000\errDU{1400}{2300}$\,km s$^{-1}$  and $ 2400^{+2900}_ {-1400}$\,km s$^{-1}$ for the  \mgxi\   and     \mgxii\ lines respectively. These  line widths  could suggest an origin closer in than the  location of the Narrow Emission Line Region, like from the outer   BLR. However, as we will discuss in \S4.1 these line widths could   be an artifact due to the   diffuse emission component, which broadens the lines  when it is dispersed by the gratings. Indeed,  as we will show later in \S 4.1, there is evidence that the emission  at these energies is spatially extended, which  causes a  degradation of the apparent spectral resolution.  
 We note also that in the case of the \mgxi\ triplet  the   prominence of the resonance component  with respect to the    forbidden line  is suggestive that the triplet is  mainly produced in a collisionally ionised plasma   rather than in  photoionised gas (\citealt{Porquet2000}; see also \S 4.1.1). This could indicate that at least part of the triplet emission  is associated with the circum-nuclear star forming regions, rather than the photoionised emission associated with the AGN NLR gas.
  
 \subsubsection{The \Sixiii\ and \Sixiv\ emission lines }  
Both  strong forbidden and  resonance  components of the \Sixiii\ triplet are detected  (see Table~\ref{tab:line1} \& \ref{tab:line2} and Fig.~\ref{fig:2048_plot}),  where   the forbidden  component appears to be marginally stronger
than the resonance line, but with large uncertainties. An emission line could be present at the energy of the intercombination line (see Fig.\ref{fig:MgXI_Si_4096_plot}, lower panel) but it is not statistically required  ($\Delta C=2$).  We detected both the H-like \Sixiv\ Ly$\alpha$ and Ly$\beta$   (see Fig.~\ref{fig:2048_plot} panel B and C) emission lines ($\Delta C=23.5$ and $\Delta C=9.4$, respectively).   As for the \mgxi\ triplet, we then fitted the spectra at a finer binning,  corresponding to  a spectral resolution of $\Delta \lambda \sim 0.01\AA $\,
 for both the MEG and the  HEG  spectra. Both the forbidden and resonance components of the He-like triplet are resolved and tying the widths of these lines to a common value results in an overall line width of $\sigma = 7^{+3}_{-2}$\, eV;  the improvement of the fit with respect to two unresolved emission   lines is $\Delta C=11$.  A similar broadening is also required  for the  \Sixiv\ Ly$\alpha$ and Ly$\beta$ emission lines  ($\sigma= 5\errUD{5}{2}$\,eV).   These  line widths would correspond to a velocity broadening of 
 $FWHM\sim 2500$\,km s$^{-1}$ and $FWHM\sim 1800$\,km s$^{-1}$, for the \Sixiii\ and \Sixiv\ lines, respectively.  However as we will discuss later, similar  to the Mg triplet, the line widths could be in part explained with the presence of  a  spatially extended component.

 \begin{table*}
 \centering
\caption{Summary of the emission lines  detected in the HETG grating data, when we considered the  spectra binned at HWHM ($0.01$\AA\ for the MEG and $0.005$\AA\, for HEG).}
 \begin{tabular}{lccccc}
\hline\hline

Line & Rest Energy$^{\mathrm{a}}$    & EW & $\sigma$ & $FWHM$   & $\Delta C^{\mathrm{b}}$\\
  & (eV) &eV&  (eV)   & (km s$^{-1}$) &  \\

\hline
\mgxi\ (r)&  $1352\errDU{2}{4}$   &  $41 \errDU{17}{22}$ &  $5.8 \errDU{2.8}{4.4}$ &$3000\errDU{1400} {2300}$  & 33.5\\
\mgxii\ Ly$\alpha$ & $1473\errDU{6}{3} $  & $28 \errDU{14}{23}$  & $5.1 \errDU{2.9}{6.1}$ &$2400\errDU{1400}{2900}$       &18.2\\
 \mgxii\ Ly$\beta$ &$1749\errDU{11}{13}$  & $15\errDU{12}{24}$ & $5.1^{\mathrm{t}} $  &  $-$ & 6.3 \\
\Sixiii\ (f)  & $1838 \errDU{3}{3}$ &$ 50 \errDU{15}{20}$   &   $6.6\errDU{2.2}{3.4}$ &$2500\errDU{900}{1300}$& 31.9  \\ 
\Sixiii\ (r)  &$1864\errDU{6}{5}$ & $ 34 \errDU{16}{24}$   & $6.6^{\mathrm{t}} $  &  $-$ & 64.6\\
\Sixiv\ Ly$\alpha$ & $ 2006 \errDU{3}{4}$   & $ 41 \errDU{16}{22}$ &$5.2\errDU{2.2}{4.7}$ &$1800\errDU{800}{1700}$  &28\\
\Sixiv\ Ly$\beta$ &$2382\errDU{5}{4}$ & $ 37 \errDU{20}{29}$  & $5.2^{\mathrm{t}} $   & $-$ &11.6\\
\Sxv\ (f) & $2430 \errDU{5}{4}$ & $31\errDU{19}{26}$ & $  1^{\mathrm{f}} $  & $-$  & 12.0 \\
\Sxv\ (r) & 2469$ \errDU{3}{3}$ & $17\errDU{11}{17}$ & $  1^{\mathrm{f}} $  & $-$  & 9.6 \\
\Sxvi\ Ly$\alpha$  &$ 2622 \errDU{5}{10} $& $30\errDU{22}{34}$ &$  1^{\mathrm{f}} $ & $-$  & 5.2 \\
\Sxvi\  Ly$\beta$   & $3107 \errDU{6}{4}$ & $31\errDU{21}{29}$& $  1^{\mathrm{f}} $ & $-$  & 7.9 \\
 \Arxviii\ Ly$\alpha$ &$ 3324 \errDU{3}{1}$ & $58\errDU{28}{36}$& $  1^{\mathrm{f}} $ & $-$  & 18.9 \\
 \feka$^{\mathrm{c}}$&$ 6401\errUD{6}{6}$ &$267\errDU{63}{74}$ &$ 14.0\errDU{7.7}{8.3}$ & $1500\errDU{800} {900}$ &101.6\\
 \hline
 \end{tabular}
\tablefoot{
\tablefoottext{a}{Measured line energy in the  \sorg\ rest frame.} \\
\tablefoottext{b}{Improvement in C-statistic upon adding the line.}\\
\tablefoottext{c}{The parameters of the \feka\ emission lines refers to the best-fit model after the inclusion of the possible Compton shoulder and the $EW$ is against the total observed continuum.}\\
\tablefoottext{f}{Denotes the parameter is fixed.}\\
\tablefoottext{t}{Indicates that the $\sigma$ is tied to the width of another transition from the same ion specie.\\}
}
\label{tab:line2}
\end{table*}

  \subsubsection{The Sulphur   and Argon emission lines }
  Although weak we detected  both the  forbidden and  resonance  components of the \Sxv\ triplet, as well as  the \Arxviii\ Ly$\alpha$ emission line.  
  The \Sxvi\ Ly$\alpha$  emission line is marginally detected ($\Delta C=5.4$) together with the possible Ly$\beta$ emission line  ($\Delta C=7.1$), for the latter however the identification is not secure as within the errors the energy centroid is compatible with the forbidden component of \Arxvii. We note however that there are no strong residuals at the corresponding energy of the \Arxvii\ resonance component ($\Delta C < 2$). All these emission lines are unresolved at the HETG spectral resolution ($\sigma <2 $\,eV).  
  We   inspected the HEG spectrum   binned at $\Delta \lambda \sim 0.01\,\AA $  to better characterize the profile of the strongest lines   and  we found that  at this finer binning only loose constraints can be derived for the widths of these emission lines, which are all consistent with being unresolved,  with upper limits  on the  $ FWHM$ of $\sim 3100$ km s$^{-1}$.  
  
  \subsubsection{The \feka\ emission line}
\begin{figure}
\includegraphics[angle=-90,width=0.44\textwidth]{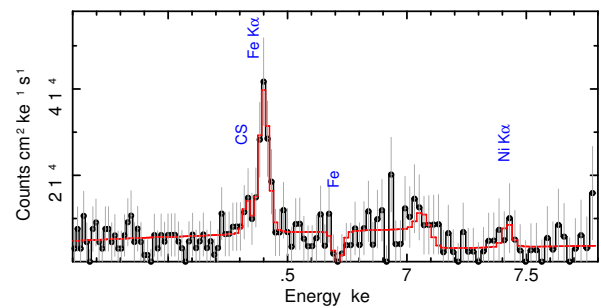}
\caption[Si_4096]{HEG spectrum binned at $5$\,m\AA\  at the energy of the \feka\ emission line.  The blue   labels mark  the main features present in the spectrum. Although it is not statistically required    we included also a Gaussian line to account for the \fekb\ emission line at $7.06$\, keV.  No clear residuals in emission  are present at the expected  energy of the \fexxv\ and \fexxvi\ lines (i. e. at  6.7 keV and 6.97 keV, respectively),  which are commonly seen in highly obscured AGN.  A clear absorption feature is detected at $\sim 6.7$\, keV, which suggests the presence of a ionized absorber.
\label{fig:Fe_4096_plot}
}
\end{figure}

 A strong \feka\ emission line is clearly detected both in the MEG and HEG spectra with a $\Delta C=112.5$ for 3 d.o.f.; the \feka\  emission  line appears to be resolved with a $FWHM= 2500^{+1000}_{-1500} $ km s$^{-1}$,  while the expected \fekb\ is only marginally detected ($\Delta C=4$). The measured $EW=290\errDU{70}{80}$ eV, or $1.2\pm 0.3$ keV with respect to the reflection component, is in agreement with the highly obscured state of \sorg\ (\citealt{Mytorus}).  As we collected  $\sim 600$ counts in the 5--8 keV energy range and   about  $\sim 90$  counts  at the energy of the \feka\ emission line we  then considered   the HEG spectrum binned at   the highest spectral resolution (i.e.  $5$ m\AA). We do not consider the MEG data as at this energy  range they have  a lower spectral resolution and  the S/N of the spectrum is too low compared to the HEG data.  The \feka\  ($E=6.399\errDU{0.007}{0.006}$ keV) emission line is resolved with $\sigma=17\errDU{9}{12}$\, eV (see Fig.~\ref{fig:Fe_4096_plot}) and  an intensity of $I=1.8\errDU{0.4}{0.5}\times 10^{-5}$\,ph cm$^{-2}$ s$^{-1}$, which is similar to the intensity  measured with all the CCD observations (see P07, B09 and Rivers et al. 2015).    
   We then tested if the line width could be in part explained with the presence of the Compton shoulder,  which may be expected when the line emission originates from scattering off Compton thick material (\citealt{Matt02,CSYaqoob}). The inclusion of a Compton shoulder only marginally improves the fit  ($\Delta C=5$) and reduces the \feka\ line width to $\sigma=14\pm 8$\,eV (or $FWHM=1500 \pm 900$ km s$^{-1}$). We note that, although loosely constrained  both  the centroid energy ($E=6.32\pm 0.18$ keV)  and intensity  ($\sim15$\% of the \feka\  intensity) of this component are in agreement with the expected values  (\citealt{Matt02,CSYaqoob}). 
   
  The main unexpected feature, which  is detected in the  HEG spectrum, is an absorption line ($\Delta C=13.8$ for 2 d.o.f.), which  is centered at $E=6.712\errDU{0.009}{0.004}$ keV with an EW of $42\errDU{15}{10}$ eV. 
 Although   weak, its energy is, within the errors,  consistent with the rest frame energy of  the $1s-2p$ resonance absorption line from \fexxv \ ($E_{\mathrm{lab}}=6.7$ keV),  thus a possible origin of the line may be  from a    highly ionized  absorber.   We then explored its properties    by  replacing  the  Gaussian absorption line with a fully covering ionized absorber  modeled with an \textsc{xstar} (\citealt{xstar}) absorption grid that  covers a wide range  in ionization (log $\xi=0-6$ erg cm s$^{-1}$) and $\nhsym$ ($1\times 10^{18}-10^{24}$ \nh). As the line is narrow we tested a  grid  that     assumes a turbulence velocity of $\sigma = 300$ km s$^{-1}$ and  found that the absorption line is   well reproduced by an  absorber with $\nhsym \sim  10^{23}$\,\nh and log $\xi \sim 2.8$\,\logxi\ with no  net outflow velocity  ($v_\mathrm{out}<1500$ km s$^{-1}$).  This is  consistent  with  an absorber located at  distances larger than the typical disk winds (\citealt{Gofford2015,Tombesi2012,Tombesi2013,Nardini2015}; see  also \S 4.2). 
  
  \section{Discussion}
The high resolution  X-ray spectrum  of  the Compton thick state  of \sorg\  is characterized by an   extremely low continuum level and is  rich in 
  emission lines    up to the  Fe K band, which, as is  seen  in   other  highly obscured Seyferts,  could  originate in a  photoionised gas that can be identified with the NLR gas (\citealt{Sako00, Kinka02, Guainazzi07, Kallman1068}).  Commonly used diagnostics for the emitting regions, when high resolution spectra are available, are: the ratio between the  forbidden and resonance components of the He-like triplets (\citealt{Porquet2000}), the  detection of the radiative recombination continuum  (RRC) transitions,  the relative strength of the higher order  transitions  (\citealt{Kinka02}) and the broadening of the line profiles.   From these diagnostics we could in principle infer a first order estimate of the density, temperature (or ionization) and location of the emitting region  in \sorg, but    this technique   requires a high signal to noise spectrum, in order to place stringent constraints on the line profiles and intensities. We also note  that the interpretation of the line widths and ratios could be complicated if the emitting  region  is extended or the emission originates from a hybrid of a photoionised and  collisionally ionized plasma. In the former case, the measured line widths could be artificially broadened as a result of the spatial extent of the emission, while the latter may result in a mixture  of photoionized and collisionally ionized gas contributing to the forbidden and resonance lines of  the He-like triplets. 
  However a  complementary  and important aid  comes from the  analysis of the high spatial resolution  maps of the X-ray emission lines (like the  He- and H-like O, Ne, Mg and Si)   that can be obtained with \chandra.  
  This latter technique has been exploited for several nearby AGN,  for which  the narrow band  \chandra\  images  and their comparison with the optical and infrared images collected with HST  (\citealt{Bianchi06,Wang2011})   made it possible  to map the location of the X-ray absorbers and/or  emitters on size-scale of a few tens/hundreds of pc (e.g. NGC~4945, \citealt{Marinucci4945}; NGC~1365, \citealt{Wang09}, NGC~3393 \citealt{Maksym3393}; NGC~7582, \citealt{Bianchi07}). In particular these studies showed a correlation between the spatial distribution of the [O\,\textsc{iii}]\, $\lambda5007$ emission   and the soft X-ray maps, strengthening the hypothesis  that the same photoionized gas  is responsible for both the optical NLR and   the  soft X-ray emission lines.\\
   
  As shown by B07, from the analysis of the  archival (snapshot) ACIS-S imaging observations,   the soft X-ray emission  of NGC\,7582 is extended   and presents a complex morphology, which varies in different energy bands,   possibly tracing a stratification in the ionization of the emitting region  as well as the presence of a  Compton thin absorber located at   radial scales larger than the putative torus.  By comparing the   HST   and soft X-ray images  it was also inferred that  this latter absorber is most likely associated with  the   dust lane  of \sorg. Indeed   the softer X-ray emission comes mostly from  the  regions  of the galaxy that are less affected by the dust absorption. We note that  B07 already showed that the   soft X-ray emission is  more extended than the starburst region and  highly inhomogeneous   (see Fig.~2 and Fig.~3 of B07)  with  two clear hotspots that  cannot be    directly associated   with the location  of the strongest star forming regions. This suggested that  the  origin of the spatially extended soft X-ray emission   cannot be solely ascribed to a collisionally ionized gas and  requires a medium photoionized by the central AGN. 
   \begin{figure*}
\centering
\includegraphics[angle=0,width=0.42\textwidth]{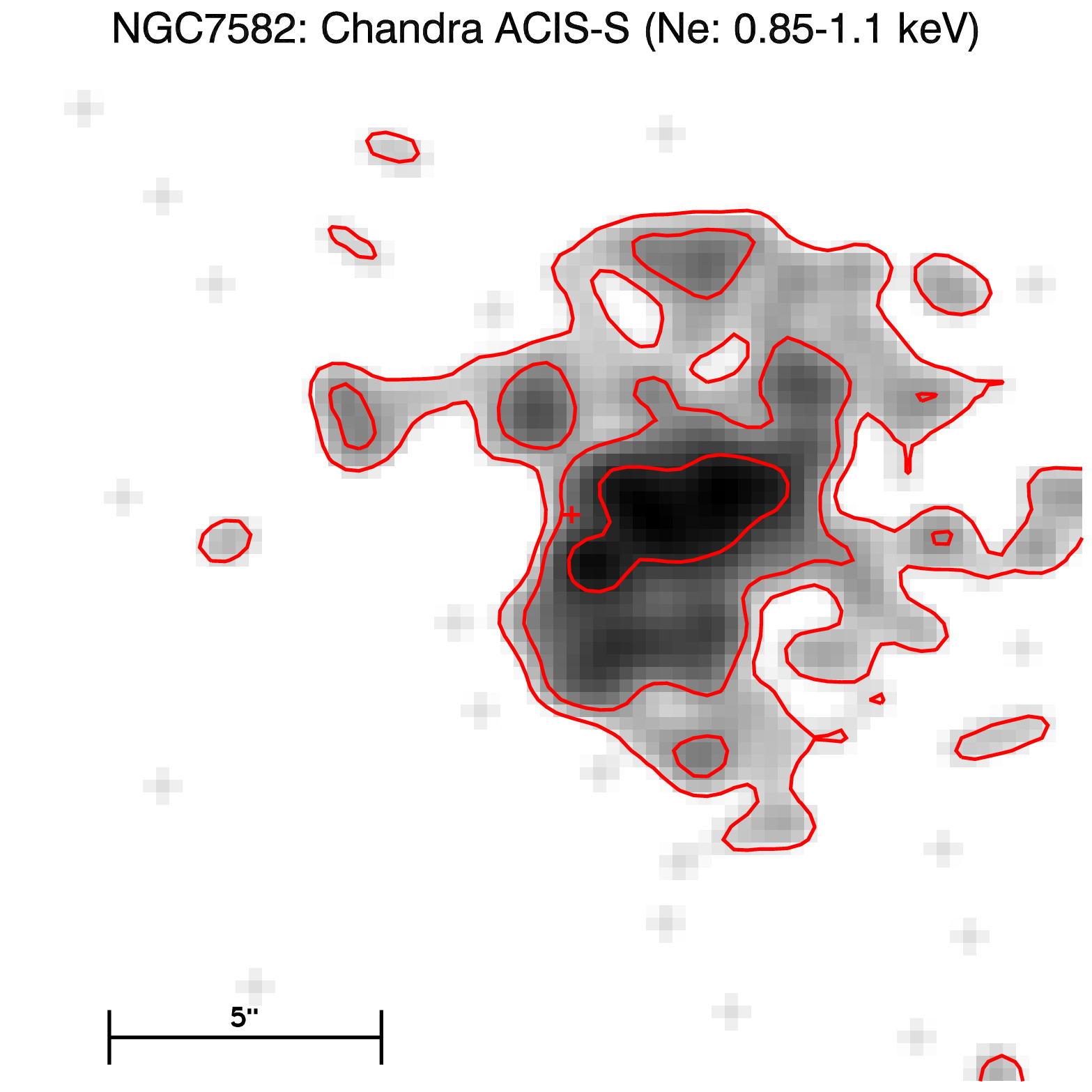}
\includegraphics[angle=0,width=0.42\textwidth]{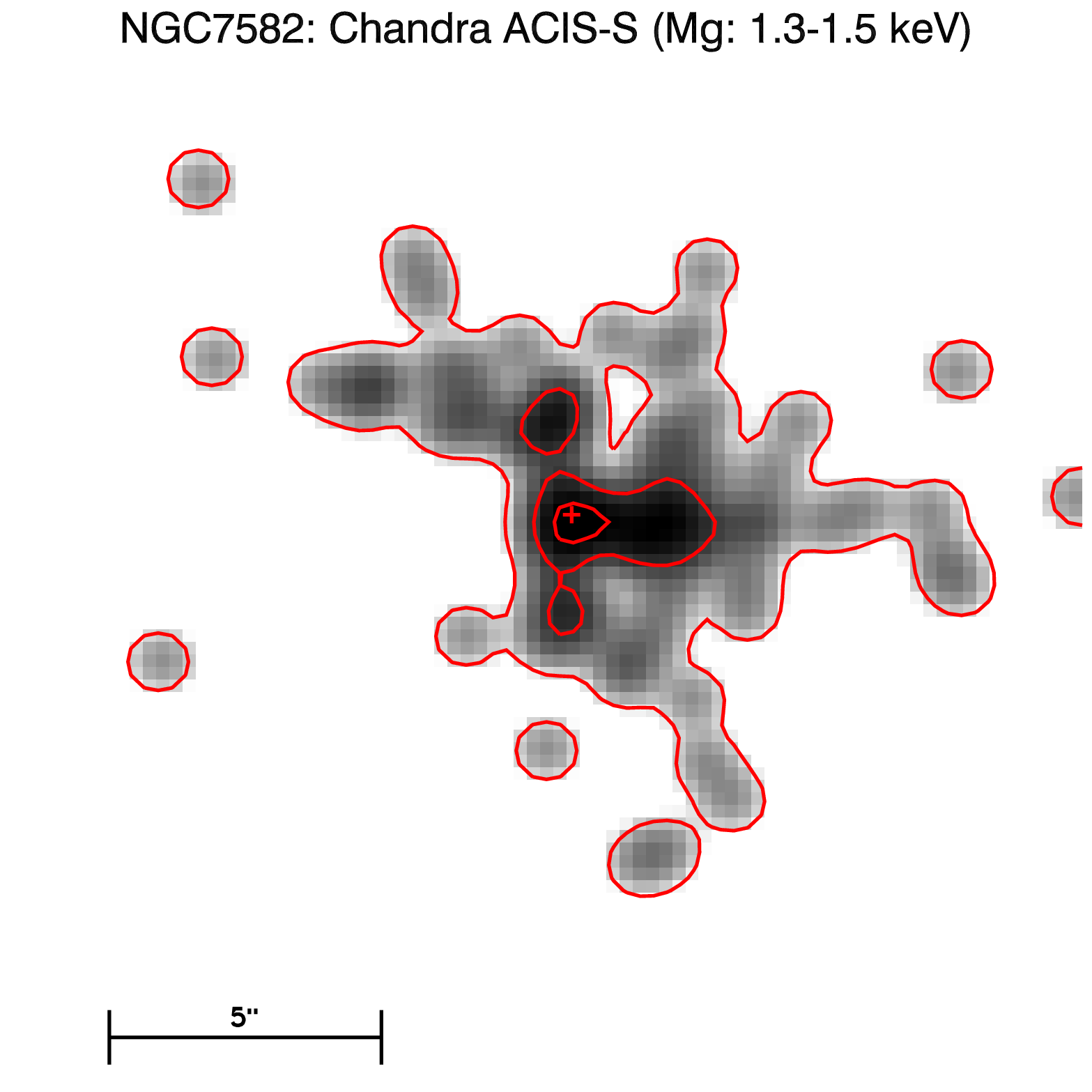}
\includegraphics[angle=0,width=0.42\textwidth]{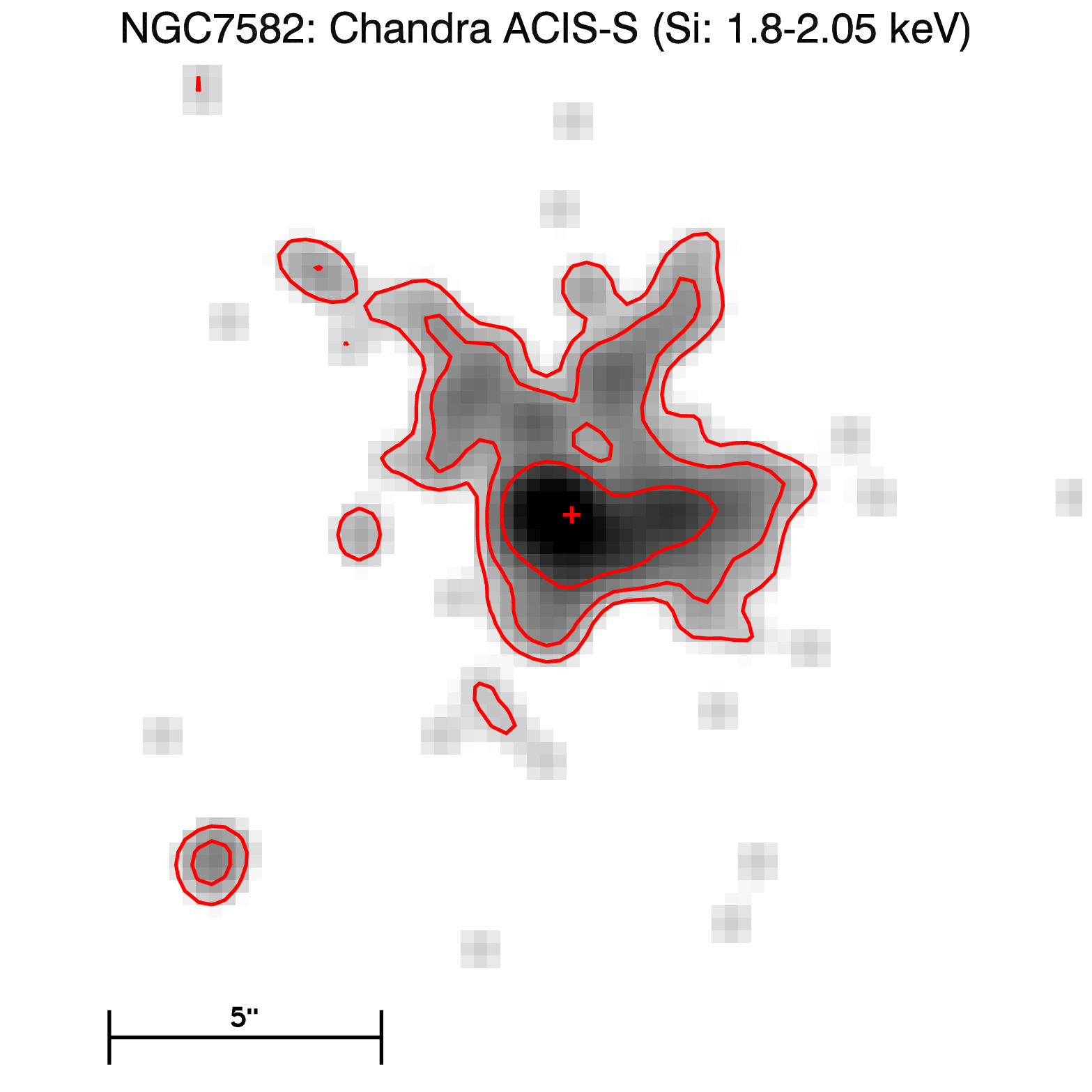}
\includegraphics[angle=0,width=0.42\textwidth]{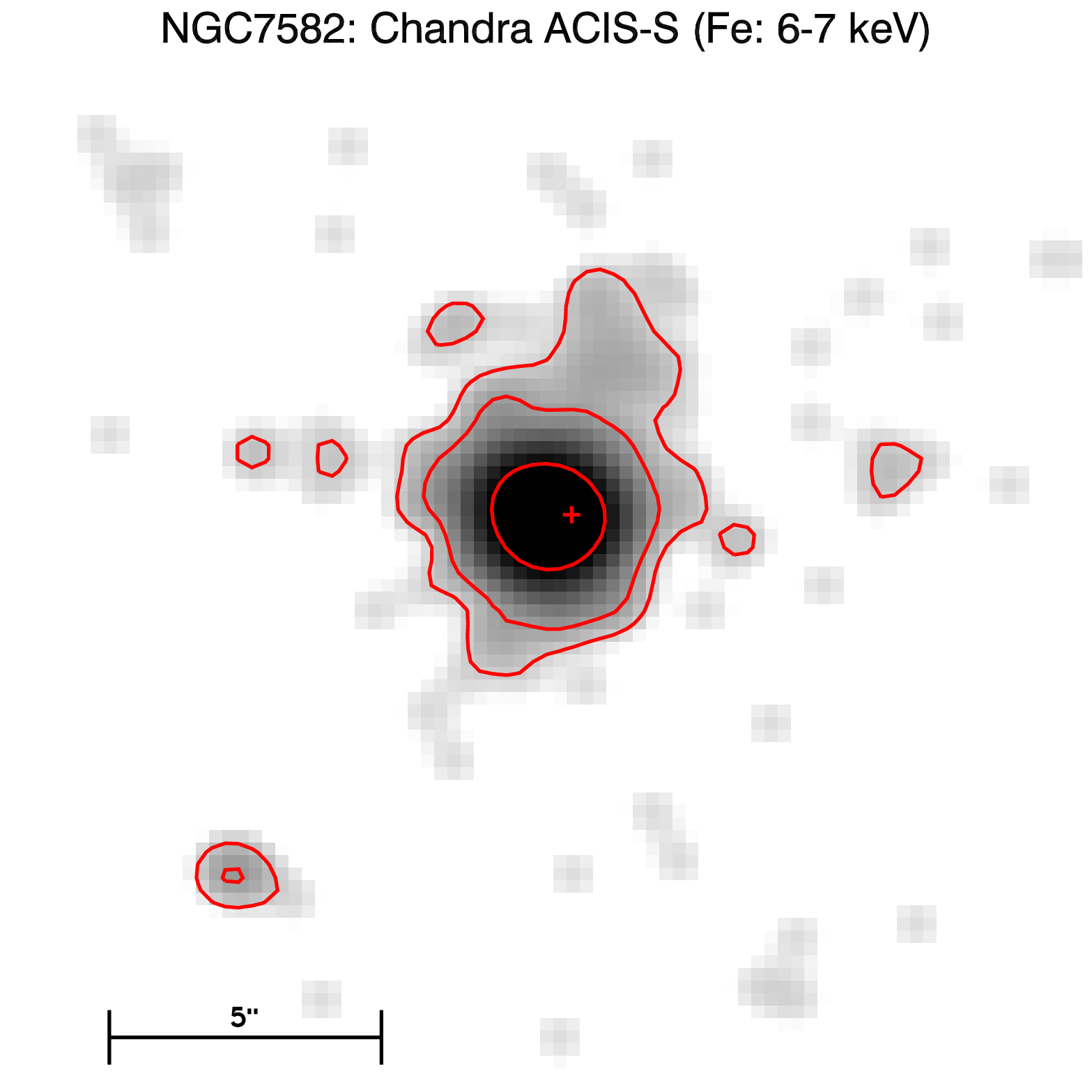}
\caption{ACIS-S zeroth order images of NGC~7582 mapping,  from top to bottom, the   Ne\,\textsc{ix-x} (upper left panel), the   Mg\,\textsc{xi-xii}   (upper right panel),  the Si\,\textsc{xiii-xiv} (lower left panel) and  the \feka\ (lower right panel) energy ranges. In all the images the red crosses mark the coordinates of the  nucleus (from NED) and the spatial scale is reported at the bottom (note that for NGC\,7582 $1''$ correspond to 109 pc). The emission appears to be spatially extended  in almost all the energy ranges, with a different morphology   depending on the energy.   The emission is progressively less extended with increasing energy and at the \feka\  emission line it is almost consistent with a point like source  albeit a weak elongation in the north direction.  We note that the \chandra\  PSF HEW is 0.5$''$.
\label{images}
}
\end{figure*}
\subsection{The origin of the soft X-ray emission line component in \sorg}
The properties of the emission lines that we detected in the      HETG spectra of \sorg\,     confirm at first order the scenario derived from the analysis of the RGS data (P07), where the    detection of the strong forbidden components in the He-like triplets  together with the  O\textsc{VIII}  RRC suggested a strong contribution from a photoionised plasma. However,  the 
presence of both the forbidden and resonance components in the He-like triplets may indicate a possible hybrid  of a   collisional and photoionised plasma, making any line diagnostic ratios hard to interpret  (i.e.  in terms of density and temperature, see \citealt{Porquet2000}). Furthermore, the spatial extent of the soft X-ray emission of NGC 7582  (see B07) complicates the physical interpretation of the soft X-ray line widths in terms of  their location of origin  from the black hole. Therefore we investigated the spatial distribution  of the soft X-ray emission, which  when coupled with the line profiles could in principle allow us to infer the  nature of the emitting gas. Thanks to the longer exposure   of the new observation, with respect to the previous  20 ksec ACIS-S exposure,   and  the availability of the grating data,  we can now create maps at higher energies  than the ones investigated by B07 and compare them directly to the line emission observed in the high  resolution spectra.\\

To explore this, we  selected four  relatively narrow bands centered on the most prominent emission line complexes; our  main aim is   to investigate if the spatial distribution of the emitting gas  could explain some of the broad profiles that we see in the spectrum.  For the analysis we considered the longest (173 ksec) of the current ACIS-S/HETG exposures and used the zeroth order to extract the images. We  extracted  four images of the central region  in the following bands: $0.85-1.1$\, keV (\neix-\nex), $1.3-1.5$\, keV (\mgxi-\mgxii), $1.8-2.05$\,keV(\Sixiii-\Sixiv) and $6-7$\,keV (\feka).  The four adaptively smoothed maps  of the central  $20''\times20''$ regions are  shown in Fig~\ref{images}, where we  overlay the  contours to highlight the spatial distribution of the X-ray counts and mark the expected position of the nucleus as reported in NED (NASA/IPAC Extragalactic Database\footnote{http://ned.ipac.caltech.edu/}).   In all the images the X-ray emission of \sorg\   appears to be extended  over different size scales depending on the energy band. Even without extracting a  radial profile,  it is clear that  the  X-ray emission is extended 
 beyond the \chandra\ PSF,  whose  on-axis Half Energy Width (HEW) is $0.5''$, as it is asymmetric in almost all the energy bands.   Only the Fe-K band image is more symmetric and dominated by the central point-like source.  

 \begin{figure*}
\includegraphics[angle=0,width=0.5\textwidth]{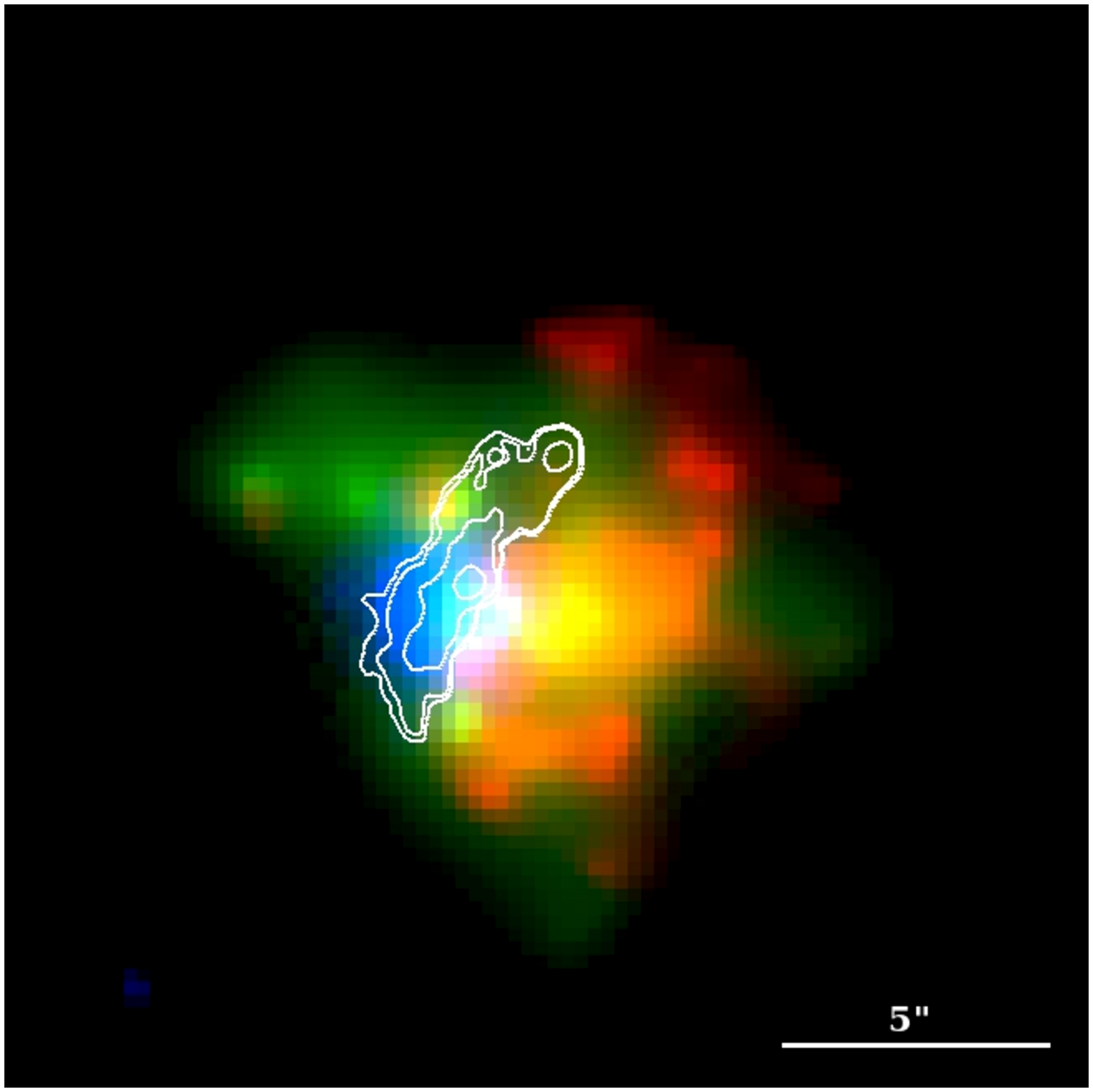}
\includegraphics[angle=0,width=0.5\textwidth]{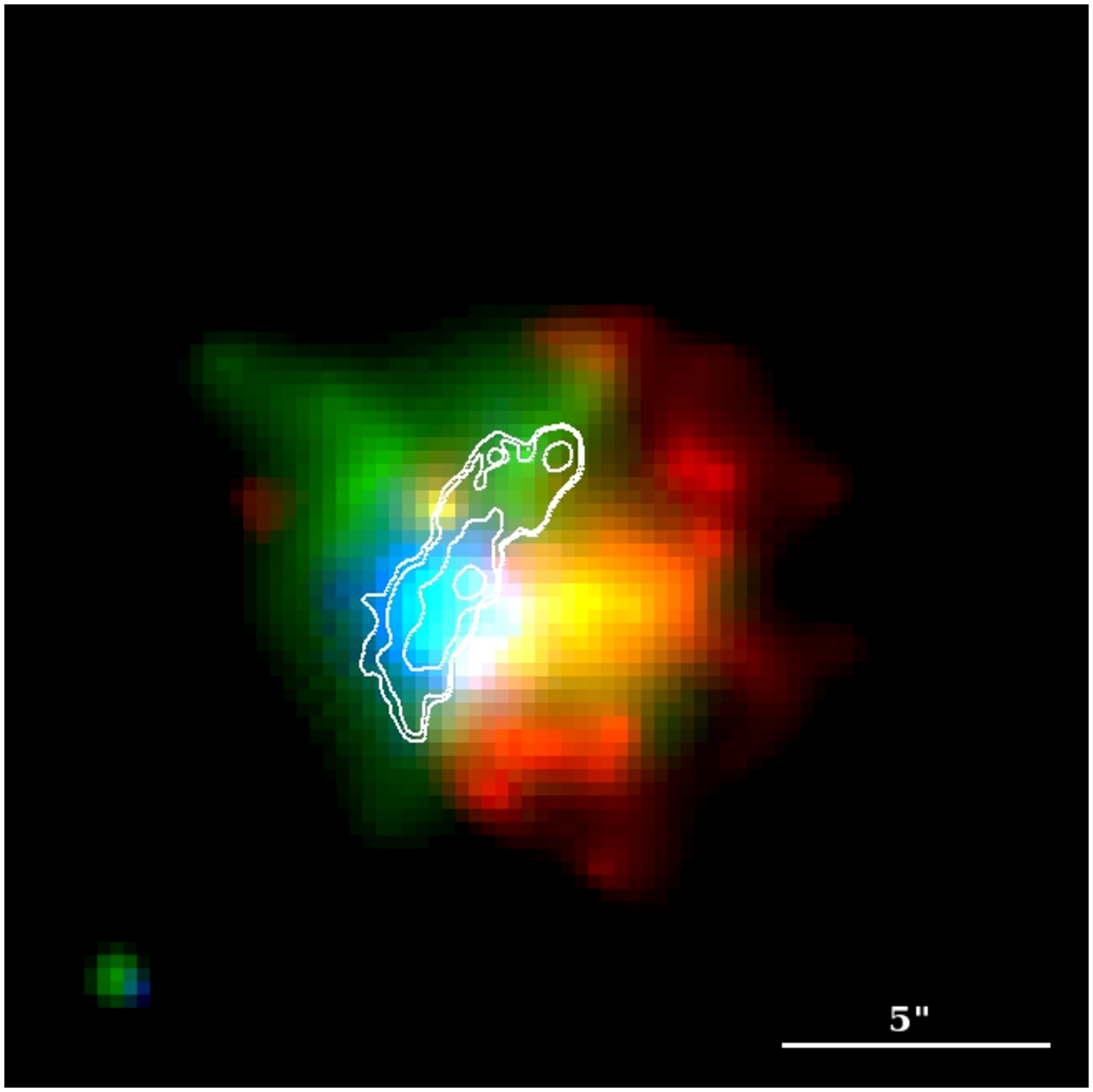}
\caption[Si_jpeg]{False color images of the central region ($20''\times 20''$) of NGC~7582.  Left panel: the red represents the Ne  band, the green is  for the Mg and the  blue represents the harder emission (6--7 keV).  In both  of the panels the white contours show the location of the dust lane as derived from the HST images. The spatial scale is reported at the bottom (note that the \chandra\  PSF HEW is 0.5$''$). The Fe K band emission is less extended and close to a point like appearance, while both the Mg and Ne emission are clearly  extended. We also note that  the Mg diffuse emission is more  on the east side of the hard nucleus  with respect to the Ne, which although more  extended is  mainly  distributed  on the west side.  This is due to the effect of obscuration by the dust lane present in NGC\,7582, which affects the lower energy band more strongly. Right panel:  the red and blue are again the Ne and Fe bands  while  the green is  for the Si band.  The narrower  source size  in the Si  band   with respect to the Ne  and Mg emission is  apparent as well as the residual excess with respect to the Fe emission. 
\label{fig:Mg_Si_jpeg}
}
\end{figure*}

 In  Fig.~\ref{fig:Mg_Si_jpeg} we show   color  composites    of the  adaptively smoothed  images  in different energy bands  corresponding to  Ne (red),   Mg  (green, left panel) or Si (green, right panel) and Fe (blue).     These false color maps demonstrate that  the Ne emission  is  more extended with respect to the Si (Mg)  and Fe bands.  Furthermore, we confirm that the peak of the emission is  offset  between the softer and harder images (\citealt{Dong}); while in the Fe-K band image the peak is almost centered on the nucleus, the Ne band emission is offset by $\sim 1.5-2 ''$, which corresponds to a projected distance of $\sim 200$\,pc (for   \sorg\  $1''$ corresponds $\sim 109$\,pc).   As already noted by B07 most of the softer (i.e. around the  Ne band) X-ray emission comes from the west side of the galaxy  and its asymmetric distribution can be explained with the presence of the dust lane which  is in the north east direction.   Diffuse  X-ray emission on the west side of the hard X-ray nucleus   appears in  the Mg and Si energy ranges (see Fig~\ref{images} and Fig.~\ref{fig:Mg_Si_jpeg}).   This is in agreement with the scenario  proposed by  B07,   where  the   dust lane of \sorg\  can  obscure the X-ray emission  below 1 keV,  but it  becomes   thinner  at higher energies.  We note that the wide extension ($\sim 6$ arcsec) of the Ne band emission  (Fig~\ref{images}, upper left panel) as well as the offset with respect to the  nucleus, and thus the center of the grating extraction region,  can  possibly explain the  non detection of the  \neix\ and \nex\ emission lines in the spectra as well as the low photon statistics below $\sim 1$\,\, keV.    \\
 
\subsubsection{The Mg and Si emitter}
From the modeling of  the  Mg emission  lines (\S~3.2.1) we noted  two main results:  the lack of  a significant detection of the forbidden component of the He-like triplet and a broadening of all the emission lines  below 2 keV with a width   of about $\sim 5 $\, eV. This latter result   could explain why a thermal emission model fails to account for the soft X-ray lines, as the emission lines in the \textsc{MEKAL} model are  narrower than  those observed.  The  lack of a forbidden emission line  could be interpreted in a scenario where the emission originates mainly in a collisionally ionized plasma or in high density hybrid gas  ($n_e>10^{14}$\,cm$^{-3}$)  where the forbidden component is collisionally suppressed. However, taken at face value  a broadening of $ \sim 3000$ km s$^{-1}$, if interpreted  as  due to virial motion,  would place this emitting gas  at a distance of about $\sim 0.04$ pc ($M_{\mathrm{BH}}\sim 5.5\times   10^7$ M$_\odot$; \citealt{Mass_ref}) corresponding to the  outer BLR, which is at odds with the extended nature of the  soft X-ray emission.  To reconcile  these results we note that the  extension    seen in the   Mg\,\textsc{xi-xii} map  over $\sim 6 ''$  (see Fig~\ref{images}, upper right panel) can    explain the line profiles detected in the MEG spectrum, as one of the main consequences of   the diffuse emission is a degradation of the spectral resolution of the grating spectra\footnote{See Fig.~8.23 of the \chandra\ Proposer's Observatory Guide;  http://cxc.harvard.edu/proposer/POG/}. Assuming  a  source size   of $\sim 5''$,  the spectral resolution of the MEG at the energy of the \mgxi\  and \mgxii,  is degraded  down to $E/\Delta E\sim 30-40$ (i.e. by  a factor of 10 with respect to a point like source), which  corresponds to an apparent  line broadening of $\sigma\sim 15$\, eV.  This can easily account for the line widths observed in the Mg band. Thus the widths of   the  \mgxi\   and \mgxii\ emission lines  (see Table 2) can not be directly  interpreted as due to  a velocity  broadening as they are most likely an artifact of the spatial broadening.   We thus conclude that a strong contribution to the      Mg emission  lines   is  from a diffuse and collisionally ionized gas, as this could explain the strong resonance line from \mgxi\ and the relative weakness of the forbidden component.

  Our spectral deconvolution   of    the  \Sixiii\ and \Sixiv\  He-like triplets revealed that the emission lines   are again resolved with a similar $FWHM\sim 2000$ km s$^{-1}$ and   with a slightly  stronger contribution    from the forbidden  with respect to the resonant component   (see Table~\ref{tab:line1}  and \ref{tab:line2}).  This  suggests  that we are dealing with  an hybrid plasma,  where the density is less than the critical value of $n_e<10 ^{16}$\,cm$^{-3}$, for which  the  forbidden component  would be collisionally suppressed.   Furthermore, even if the large uncertainties   on  the lines intensities do not allow us to derive more stringent constraints,   the  detection  of the strong forbidden component  suggests a   contribution  from a photoionised    gas to the  forbidden emission. The  map for the  Si  band shows  that the emission is  slightly less extended  (over $3 ''$) and    more centered on the nucleus and also more symmetric than the Ne and Mg emission.  However,    also in this case the  source size corresponds   to  an apparent  line broadening  in the MEG and HEG data similar to the measured widths. 
 
\begin{figure}
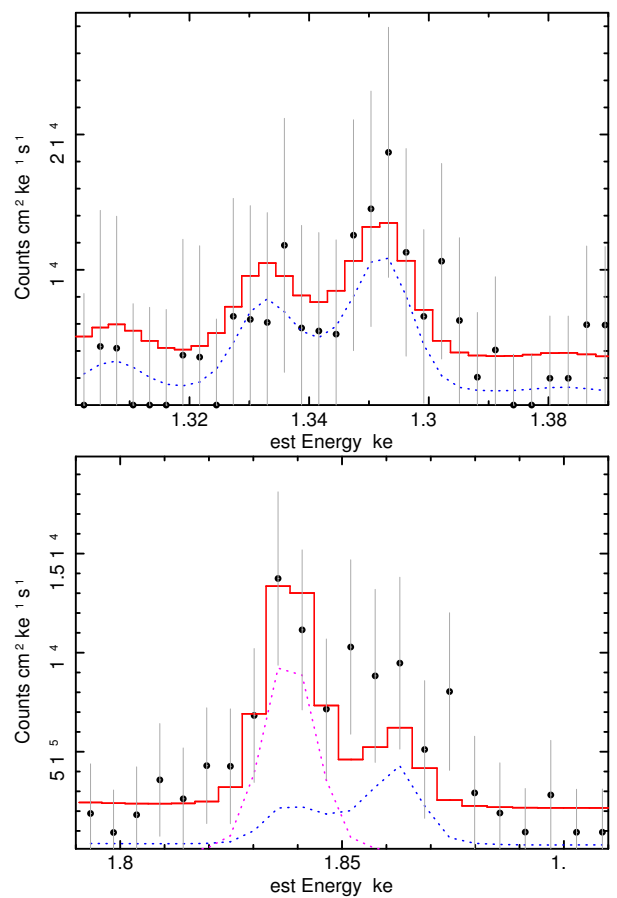

\begin{center}
\includegraphics[angle=-90,width=0.45\textwidth]{Mg_mekal.eps}
\includegraphics[angle=-90,width=0.45\textwidth]{new_Si_mek.eps}
\end{center}
\caption{Zoom   into the MEG spectrum  (at HWHM resolution) at the energies of the \mgxi\  and \Sixiii\ triplets when modeled with the addition of a thermal emission model (blue dotted line).     The upper panel shows the \mgxi\ region, where  the  \textsc{mekal} component can reproduce all the profile with no need of an additional Gaussian emission line component.  The  lower panel  shows  that in order to reproduce the  \Sixiii\ triplet, we need to include  a strong emission line component (magenta line) to account for the  residual emission at the energy of the  forbidden line.   In both the panels the red  line shows the  best-fit model where we included a  \textsc{mekal} component (see \S 4.1.1).
\label{mekal_plots}
}
\end{figure}

 We then attempted to quantify how much of the line emission may originate from a collisionally ionized plasma. In particular we tested whether it was possible to  account for most of the \mgxi\ and  \Sixiii\ triplets with a collisionally ionized  emission  model, thus we replaced these  Gaussian   lines with a  \textit{MEKAL} component. We  accounted for the measured line widths   of  $\sim 5$ eV  by convolving the thermal line emission with a Gaussian broadening (via the  \textit{gsmooth}  model in xspec). Although  this model is statistically  worse   ($\Delta C=21$  for 6 d.o.f) with respect to the  Gaussian emission line best-fit model,  it can  formally account for the \mgxi\  triplet, while for the \Sixiii\  complex we still need a strong contribution from a photoionised gas, indeed we still require an additional strong forbidden line (see Fig.~\ref{mekal_plots}).   We note that the   derived  \textsc{mekal} temperature   is typical of a starburst emission ($kT=0.57\errDU{0.07}{0.09}$ keV).   Its    flux    ($F_\mathrm{0.5-2\,keV}\sim 2.7\times 10^{-13}$\, \flux)     is  in agreement with the  flux of the diffuse component measured in the previous \chandra\ imaging observations after the removal of the central nuclear source (\citealt{LaMassa2012}). The soft X-ray luminosity  ($L_\mathrm{0.5-2\,keV}\sim 1.7\times 10^{40}$\, \lum)   is also  in agreement with the expectation from  the  infrared luminosity of \sorg\  ($L_\mathrm{FIR}\sim 9.6\times 10^{43}$\,\lum, \citealt{Taniguchi1998}) and  the  $L_\mathrm{X}-SFR$ relations derived from the X-ray observations of several starburst galaxies      (\citealt{Ranalli2003,Mineo2012,Pereira-Santaella2011}). In particular,  assuming that most of the    infrared luminosity  can   be ascribed to the star formation activity,  a $L_\mathrm{FIR}\sim9.6\times 10^{43}$\,\lum\ implies  a star formation rate  (SFR) of about $\sim 4 M_\odot$/yr (\citealt{Kennicutt1998})  for which the   $L_\mathrm{X}-SFR$  relations  would then predict a soft X-ray luminosity of  $1-2\times 10^{40}$\, \lum.   
 
 \subsubsection{The location and density of the emitting gas}
From the normalization  of the thermal  component  ($A_\mathrm{MEKAL}=1.74\errDU{0.47}{0.53}\times 10^{-4}$\,cm$^{-5}$) we can  now derive a  first order estimate of the plasma density ($n_\mathrm{e}$). Indeed,  by definition,  the normalization of  the {\sc MEKAL}  model   is  proportional to the volume emission measure ($E.M. = \int n_e n_H dV$) of the X-ray emitting gas with $A_\mathrm{MEKAL} = 10^{-14} E.M. /[4\pi D_{A}^2 \times (1 + z)^2] $ cm$^{-5}$, where $D_\mathrm{A}$ is the angular-diameter distance to the source and $z$ is the redshift.  We assumed that  the gas  extends for $\sim 500$\, pc (as derived from the soft X-ray  images)  and derived a density of    $n_\mathrm{e}\sim 0.3 $\,cm$^{-3}$ for a uniform spherically symmetric  gas distribution. For  an inhomogeneous gas with emitting clumps with a  reasonable size scale of $\Delta R/R=0.1$ (corresponding to a   volumetric filling factor of $f_\mathrm{v}\sim10^{-3}$) the density will be higher and of order of  $n_\mathrm{e}\sim 10$\,cm$^{-3}$.  
 
\begin{figure}
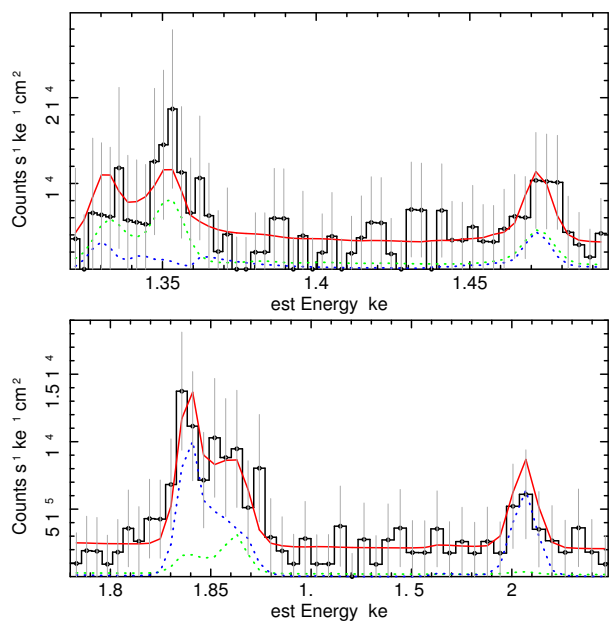

\begin{center}
\includegraphics[angle=-90,width=0.45\textwidth]{meg_Mg_xstar2.eps}
\includegraphics[angle=-90,width=0.45\textwidth]{meg_Si_xstar2.eps}
\end{center}
\caption{Zoom   into the MEG spectrum  (at HWHM resolution) at the energy of the \mgxi\  and \Sixiii\ triplets when the emission component is  modeled with  a thermal emission model (green dotted line) and a photoionized emitter, modeled with  an \textsc{xstar} grid (blue dotted line).     The upper panel shows the \mgxi\ region, where  the  \textsc{mekal} component can reproduce  the profile with  a weak contribution from   the \textsc{xstar}  emission  component.  The  lower panel  shows   that    \Sixiii\ triplet as well as the \Sixiv\ Ly$\alpha$   are  mainly accounted for by the  photoionized emiter component (blue dotted line), while the  thermal emission component   accounts for the  residual emission at the energy of the  resonance emission line of the   \Sixiii\ triplet.   In both of  the panels the red  line shows the  best-fit model where we the emission line component is accounted for   by a collisional gas (\textsc{mekal}) and photoionzed  emitter (\textsc{xstar}; see \S 4.1.2).
\label{mekal_xstar_plots}
}
\end{figure}

   To derive a first order estimate of the location and density of the photoionised emitter that is  responsible for the soft X-ray emission lines, or at least the forbidden ones, we  replaced the soft X-ray Gaussian  emission lines with  an  \textsc{xstar} emission model  and then     fitted  the Mg and Si emission lines.   This model assumes a turbulence velocity of $v_\mathrm{turb}= 1000$ km s$^{-1}$ to account for the residual  broadening.  We considered only the MEG data as  in the energy range of interest ($1-2$ keV)  the HEG data have a lower S/N; the resulting best fits of the \mgxi\ and \Sixiii\ He-like triplets are shown  in Fig.~\ref{mekal_xstar_plots}.  The ionization of this photoionized emitter is log\,$\xi=2.4 \errDU{0.3}{0.4}$\,\logxi and its normalization is $k_\mathrm{xstar}=(5.5\pm 1.6)\times 10^{-5}$, for a column density of $\nhsym\sim 10^{21}$\,\nh.   The  temperature  of the thermal emission component  is again  $kT\sim0.56$ keV and its  soft X-ray  luminosity   is $L_\mathrm{0.5-2\,keV}\sim 1.2\times 10^{40}$\, \lum,  which would correspond to  approximately 60\% of the    observed soft X-ray luminosity   ($\sim 2\times 10^{40}$\,\lum). However this  latter estimate must be considered as an upper limit;   due to the lack of enough S/N below 1 keV we cannot exclude the presence in the  model of a  lower ionization  photoionized emitter,  which     could be responsible  for the  diffuse emission seen at the energy of the Ne emission lines. \\ 

In terms of the  photoionized gas  properties, a first order estimate of its density and covering factor   can be derived from the \textsc{xstar} component normalization.   For a shell of gas photoionized by a central source, the normalization  of the emission component, $k_\mathrm{xstar}$,  is defined as $k_\mathrm{xstar}=f L_{38}/D_\mathrm{kpc}^2$ (Kallman et al. 2004),  where $L_{38}$ is the ionizing luminosity over the $1-1000$ Rydberg range  in units of $10^{38}$\,\lum, $D_\mathrm{kpc}$  is the distance  to the  source  in kpc and $f$ is a geometrical factor of the emitting region with respect to a uniform shell covering a solid angle  $f=\Omega/4\pi$.   As \sorg\ was observed in a highly obscured state,  before deriving the  intrinsic luminosity we  need to  consider the Compton-down scattering effect   that can be important at this level of obscuration (\citealt{Mytorus}). We  thus  included  in the best-fit  model an additional absorber ({\sc CABS} model in XSPEC)  with the same column density  of the neutral absorber to correct for it and derived an upper limit on the  intrinsic ionizing luminosity of   $L_\mathrm{ion} \sim  7\times 10^{42}$\,\lum.  Thus  for the above ionizing luminosity    and the distance  of NGC\,7582 ($\sim 22.7$\,Mpc),  assuming a fully covering spherical shell we derive $k_\mathrm{xstar}=1.4\times 10^{-4}$. For  a given column density the  measured normalization provides then a first order estimate  of the covering factor of the  emitting gas.  Thus assuming a column density as above of $10^{21}$\nh\ the   ratio between the observed  ($k_\mathrm{observed}=(5.5\pm 1.6)\times 10^{-5}$)  and predicted normalizations gives a covering factor $f=0.4\pm0.1$. We then fixed the column density to a higher value of $10^{22}$\,\nh\ and derived   a covering   factor of only 4\%,  which is probably too low given the spatial extent of the soft X-ray emission.  Similarly a lower limit    of $\nhsym>2\times 10^{20}$\,\nh\ on the gas density was obtained by fixing the normalization of the emitting component to the predicted value  for a fully covering shell and allowing   the  $\nhsym$ to vary.  

 An estimate of the location and density   of the emitter can be  then obtained using the definition of the ionization parameter  $L_\mathrm{ion}/\xi=nR^2$; where $L_\mathrm{ion}$ is the intrinsic     ionizing luminosity over the $1-1000$ Rydberg range   and $n$ is the  gas density.   Then assuming again a typical size  of the emitting region of few  hundred pc,   we derived a gas density of about $n_e\sim 0.3$\,cm$^{-3}$.  Alternatively assuming  that the gas is not highly clumped (i. e. $\Delta R/R\sim 1$)  and the lower limit on the column density derived above, we can estimate  a distance of $R\sim 1.4\times 10^{20}$\,  cm, which is  consistent with the location of the diffuse emission seen in the X-ray images, and  a density of  order of $n_e\sim 1.4$\,cm$^{-3}$.    Although these are an  order  of magnitude estimates, the similarity of the derived  densities  suggests that  the gas is not highly clumped.
  Indeed, if the gas were too clumped  ($\Delta R/R\sim 0.1$) the density of the single clouds would be  much higher, thus   to maintain the same ionization balance, as measured with the \textsc{xstar} best-fit model, the distance would be as low as few parsec,  which is inconsistent with the X-ray images. Thus the most likely scenario is   that the emission line component originates in a high-ionization  and low-density gas ($n_e \sim 1$\,cm$^{-3}$) which is extended on scales of few hundreds pc. The column density of this gas is low and in the range of $2\times 10^{20}-10^{21}$\,\nh and has a rather high covering factor.  
A plausible scenario is thus that some of the soft X-ray emission lines are produced in star forming regions, which are embedded in a larger scale gas which is photoionised in  part by  the central  AGN and in part by the  circumnuclear  starburst itself.  A similar scenario has been proposed for other highly obscured Seyfert 2, where both  high spectral  and spatial resolution data were available (i.e. Mrk\,573, \citealt{Bianchi2010} NGC\,1365, \citealt{Wang09}; NGC\,4151, \citealt{Wang2011}). \\

\subsection{The Fe K band}      
As discussed in \S 3.2.4 the  measured  $EW\sim 1.2\pm 0.3 $ keV  (with respect  to the reflection component) of the \feka\  emission line is consistent    with the presence  of an optically thick absorber. At the energy of the Fe K band   the  emission map  is almost consistent  with a point like source,   with the  exception of a weak   elongation in the northern direction  (see Fig.~\ref{images}).  We can thus treat  the  \feka\ emission line width as intrinsic to the source and attempt to place some constraints on the location of the emitter/absorber.  We assumed that the broadening is due  to Keplerian motion of the emitter and  adopted  a factor $f=\sqrt 3/2$ to  correct the observed FWHM for the geometry of the emitting gas (\citealt{Netzer-Marziani2010}). Thus for a  $M_{\mathrm{BH}}\sim 5.5\times    10^7$ M$_\odot$   (\citealt{Mass_ref}) the measured  $FWHM= 1500 \pm 900$ km s$^{-1}$ would correspond to a distance of $R_{\mathrm{Fe \,K}\alpha} \sim 0.05-0.7$\,pc.  The lower and upper  limits on the radial distance would place the emitting gas at the distance of the outer  BLR and the pc-scale torus, respectively.    The derived limits for the Fe-K emitter     correspond to few$\times 10^4-10^5 \, R_\mathrm{g}$, which is in agreement with  the  typical location  of $\sim  3\times 10^4 \,R_\mathrm{g}$, derived for a sample of nearby Seyfert 2 which had  HETG observations   (\citealt{Shu2011}).\\

From previous X-ray observations of \sorg\  (P07; B09; \citealt{Rivers2015}) two circum-nuclear absorbers co-exist in \sorg: one   could be   identified   with  the parsec scale torus  and one is responsible for the fast $\nhsym$ variations and is most likely located at the BLR distance.  However, the measured intensity of the   \feka\ emission line  $I=(1.7\pm 0.4)\times 10^{-5}$ ph cm$^{-2}$ s$^{-1}$ is  in agreement with  the historical  value measured in the  X-ray observations of \sorg\  performed  over the last 10 years ($I\sim 2.2-2.4\times 10^{-5}$ ph cm$^{-2}$ s$^{-1}$), which show that both the Compton reflection component and the \feka\ emission line do not vary. We thus conclude  the \feka\ emission line properties are consistent  with reflection from the inner edge  of a Compton thick toroidal absorber (\citealt{Mytorus}), which is  most likely located  at  a subpc-, pc-scale distance from the central SMBH.\\

As we noted in \S 3.2.4  a weak absorption feature is detected at $\sim 6.71$ keV,  which could be associated with the presence of an ionized absorber. The energy centroid of this absorption is close to the expected energy of the \fexxv\  resonance transition and does not require any  blueshift. This suggests that the absorber is not part of a disk wind and  is located further out than the typical location of few  hundreds $R_\mathrm{g}$ of  the disk winds (\citealt{Gofford2015,Tombesi2012,Tombesi2013,Nardini2015}).   Furthermore, the detection of  this absorber  when \sorg\ is   in a Compton thick state   would   naturally argue for a location  outside    the variable   absorber located at the BLR distance. 
 We can derive an estimate of the    maximum distance of this absorber using  again the definition of the ionization parameter $L_\mathrm{ion}/\xi=nR^2$; where $L_\mathrm{ion}$ is the intrinsic     ionizing luminosity over the $1-1000$ Rydberg range    and $n$ is the  absorber density.  Thus  for the NGC\,7582 ionizing luminosity  of $L_\mathrm{ion} \sim  7\times 10^{42}$\,\lum,  assuming  that the absorber thickness  is less than its distance (i.e. $\Delta R/R<1$) and given the ionization parameter  log $\xi \sim 2.8$\,\logxi\ and $\nhsym \sim  10^{23}$\,\nh\   we  found $R\simlt10^{17}$ cm, which is  on the same scale of the   \feka\ emitting region.   At this distance and again assuming a radial extent of the absorbing clouds of $\Delta R/R\sim 1$  the gas density is $n_e=10^8$cm$^{-3}$.   Although these are all first order estimates,  the   derived density and distance are all suggestive  that this absorber is  the putative electron scattering region (\citealt{Krolik2001,Netzer2015} and references therein), which can be associated to the inner part of the  torus.  On the other hand the thick absorber could  be associated  with clumps within the pc-scale torus, which could vary on time-scale from  weeks to months.  
 
 \section{Summary  and Conclusion }
 We  presented the results of a detailed analysis of  a deep  ($\sim 200$ ksec) \chandra\ HETG  observation of \sorg,  when the AGN was  in a  prolonged highly obscured state   with   a column density in the Compton  thick regime ($\nhsym \sim 1.2\times 10^{24}$\nh). Thanks to the combination of the high spectral  and spatial resolution of the \chandra\ data we   have gained more insight  in  the structure for the circum-nuclear emitter/absorbers  of \sorg.   
 As seen in other well studied nearby Seyfert 2s and  in particular  for the prototype of the ``Changing look'' AGN NGC 1365  (\citealt{Braito2014,Nardini2015b,Rivers2015a,Risaliti2016}), it is now clear that also in \sorg\ the  absorbing/emitting circum-nuclear gas is  not a single homogeneous gas but  it is most likely  a multiphase   medium,   located at different  distances from the sub-pc BLR region  to the larger scale galactic absorber. 
 
The main results  on the different emitting/absorbing zones and their location,  as obtained   with this new observation of \sorg\,   can be summarized as follows:-
 \begin{itemize}
  \item  From the  \feka\ emission line  width and the presence in all the observations of a relatively constant reflection component, we know that a  $0.05-0.7$\,pc scale reprocessor is present. This could be associated with the inner edge of the putative pc-scale clumpy  torus.   As NGC 7582 is not always absorbed  by the Compton thick absorber, our line of sight   is not totally obscured by this reprocessor. Indeed,   in  past observations (B09) we were able to witness fast  $\nhsym$  variations due to an  innermost absorber located at the BLR distance.
  \item A highly ionized absorber, responsible for the absorption feature detected at $\sim 6.7$ keV,  is present on a similar distance to the \feka\ emitter and  could  be identified with the putative electron scattering region.
  \item Some of the strong soft X-ray emission lines, in particular the forbidden components, originate in  photoionzed  emitting gas that can    be  associated with the extended  and low density ($n_e\sim 0.3-1$ cm$^{-3}$) NLR gas.  From the Si\textsc{xiii-xiv} image we know that this  photoionized gas is extended, on  size-scales  of  $200-300$ pc.  Furthermore, a comparison with the \neix\ and \mgxi\  narrow band images shows that this gas is less extended  than the emitting regions  where the Ne  and Mg emissions originate.  
  \item An extended  thermal gas   is   present on large scales (few hundreds pc) and is probably associated with the strong   Ne and Mg emission.  This gas is required to explain the strong resonance components and  could be ionized by the circum-nuclear star formation activity.
   \item A larger scale Compton-thin absorber obscures part of the emission below 1 keV, as seen in the asymmetric map of the Ne band  emission. As already suggested by B07 this absorber is presumably associated with the galactic dust lane east of the nucleus. 
  \end{itemize}

  Although only few ``changing look" AGN and Seyfert 2 galaxies have been observed long enough with both high spatial  and spectral resolution it is now clear that ``Changing look'' AGN like NGC\,1365 and NGC\,7582  have a favorable line of sight  such as we can view all the stratification of these  absorbers/emitters.   Our results show that  future  deep grating observations of nearby ``changing look''  and Seyfert 2 AGN performed with \chandra\  will offer a unique opportunity to map all these   emitters and absorbers. This will allow us to  build up a more complete picture of the AGN environment and allow us to improve the Unification Model of AGN.

\begin{acknowledgements}
  We would like to  thank the anonymous referee for useful comments, which have improved this paper.
This paper has made use of observations obtained with  the  \chandra\  X-ray Observatory.  This research has made use of software provided by the Chandra X-ray Center (CXC) in the application packages CIAO. J.N. Reeves  and V. Braito acknowledge Chandra grant   GO4-15101X  and   support from NASA grant NNX15AF12G.   J. N. Reeves and E. Nardini acknowledge support of the  STFC consolidated grant. E. Nardini acknowledges funding from the European Union's Horizon 2020 research programme under the  Marie Skodowska-Curie grant agreement No. 664931.    This research has made use of the NASA/IPAC Extragalactic Database (NED) which is operated by the Jet Propulsion Laboratory, California Institute of Technology, under contract with the National Aeronautics and Space Administration. 
\end{acknowledgements}

\newpage


\begin{thebibliography}{}
\bibitem[Antonucci (1993)]{Antonucci} Antonucci, R.\ 1993, \araa, 31, 473 
\bibitem[Arnaud(1996)]{xspecref} Arnaud, K.~A.\ 1996, Astronomical Data Analysis Software and Systems V, 101, 17 
\bibitem[Aretxaga et al.(1999)]{Aretxaga1999} Aretxaga, I., Joguet, B., Kunth, D., Melnick, J., \& Terlevich, R.~J.\ 1999, \apjl, 519, L123 \bibitem[Bianchi et al.(2006)]{Bianchi06} Bianchi, S., Guainazzi, M., \& Chiaberge, M.\ 2006, \aap, 448, 499 
\bibitem[Bianchi et al.(2007)]{Bianchi07} Bianchi, S., Chiaberge, M., Piconcelli, E., \& Guainazzi, M.\ 2007, \mnras, 374, 697 
\bibitem[Bianchi et al.(2009)]{Bianchi09} Bianchi, S., Piconcelli, E., Chiaberge, M., et al.\ 2009, \apj, 695, 781 
\bibitem[Bianchi et al.(2010)]{Bianchi2010} Bianchi, S., Chiaberge, M., Evans, D.~A., et al.\ 2010, \mnras, 405, 553
 \bibitem[Bianchi et al.(2012)]{Bianchi2012} Bianchi, S., Maiolino,  R., \& Risaliti, G.\ 2012, Advances in Astronomy, 2012,  
\bibitem[Boldt(1987)]{Boldt} Boldt, E.\ 1987, \physrep, 146, 215 
\bibitem[Braito et al.(2014)]{Braito2014} Braito, V., Reeves, J.~N., Gofford, J., et al.\ 2014, \apj, 795, 87 
\bibitem[Dickey \& Lockman(1990)]{Dickey} Dickey, J.~M., \& Lockman, F.~J.\ 1990, \araa, 28, 215 
\bibitem[Dong et al.(2004)]{Dong} Dong, H., Xue, S.-J., Li, C., \& Cheng, F.-Z.\ 2004, \cjaa, 4, 427 
\bibitem[Fruscione et al.(2006)]{Ciaoref} Fruscione, A., et al.\ 2006, \procspie, 6270
\bibitem[Garmire et al.(2003)]{Garmire2003} Garmire, G.~P., Bautz, M.~W., Ford, P.~G., Nousek, J.~A., \& Ricker, G.~R., Jr.\ 2003, \procspie, 4851, 28 
\bibitem[Gilli et al.(2000)]{Gilli2000} Gilli, R., Maiolino, R., Marconi, A., et al.\ 2000, \aap, 355, 485 
\bibitem[Gruber et al.(1999)]{Gruber} Gruber, D.~E., Matteson, J.~L., Peterson, L.~E., \& Jung, G.~V.\ 1999, \apj, 520, 124 
\bibitem[Gofford et al.(2015)]{Gofford2015} Gofford, J., Reeves, J.~N., McLaughlin, D.~E., et al.\ 2015, \mnras, 451, 4169 
\bibitem[Green et al.(1993)]{Green1993} Green, A.~R., McHardy, I.~M., \& Lehto, H.~J.\ 1993, \mnras, 265, 664 
\bibitem[Guainazzi et al.(1998)]{Guainazzi98} Guainazzi, M., Nicastro, F., Fiore, F., et al.\ 1998, \mnras, 301, L1 
\bibitem[Guainazzi et al.(2002)]{Guainazzi2002} Guainazzi, M., Matt, G., Fiore, F., \& Perola, G.~C.\ 2002, \aap, 388, 787 
\bibitem[Guainazzi \& Bianchi(2007)]{Guainazzi07} Guainazzi, M., \& Bianchi, S.\ 2007, \mnras, 374, 1290
\bibitem[Kallman et al.(2004)]{xstar} Kallman, T.~R., Palmeri, P., Bautista, M.~A., Mendoza, C., \& Krolik, J.~H.\ 2004, \apjs, 155, 675 
\bibitem[Kallman et al.(2014)]{Kallman1068} Kallman, T., Evans, D.~A., Marshall, H., et al.\ 2014, \apj, 780, 121 
\bibitem[Kennicutt(1998)]{Kennicutt1998} Kennicutt, R.~C., Jr.\ 1998, \apj, 498, 541
 \bibitem[Kinkhabwala et al.(2002)]{Kinka02} Kinkhabwala, A., et al.\ 2002, \apj, 575, 732 
\bibitem[Kokubun et al. (2007)]{kokubun}Kokubun, M., et al.\ 2007, \pasj, 59, 53 
\bibitem[Koyama et al.(2007)]{Koyama07} Koyama, K., et al.\ 2007, \pasj, 59, 23 
\bibitem[Krolik \& Kriss(2001)]{Krolik2001} Krolik, J.~H., \& Kriss, G.~A.\ 2001, \apj, 561, 684
\bibitem[LaMassa et al.(2012)]{LaMassa2012} LaMassa, S.~M., Heckman, T.~M., \& Ptak, A.\ 2012, \apj, 758, 82 
\bibitem[Magdziarz \& Zdziarski(1995)]{pexrav} Magdziarz, P., \& Zdziarski, A.~A.\ 1995, \mnras, 273, 837 
\bibitem[Maksym et al.(2016)]{Maksym3393}  Maksym, W.~P., Fabbiano, G., Elvis, M., et al.\ 2016, \apj, 829, 46 
\bibitem[Maiolino et  al.(2010)]{Maiolino2010} Maiolino, R., Risaliti, G., Salvati, M., et al.\ 2010, \aap, 517, A47 
\bibitem[Markert et al.(1994)]{Markert94} Markert, T.~H., Canizares, C.~R., Dewey, D., McGuirk, M., Pak, C.~S., \& Schattenburg,  M.~L.\ 1994, \procspie, 2280, 168 
\bibitem[Marinucci et al.(2012)]{Marinucci4945} Marinucci, A., Risaliti, G., Wang, J., et al.\ 2012, \mnras, 423, L6 
\bibitem[Markowitz et al.(2014)]{Markowitz2014} Markowitz, A.~G., Krumpe, M., \& Nikutta, R.\ 2014, \mnras, 439, 1403 
\bibitem[Masini et al.(2016)]{Masini2016} Masini, A., Comastri, A., Puccetti, S., et al.\ 2016, arXiv:1609.00374 
 \bibitem[Matt(2002)]{Matt02} Matt, G.\ 2002, \mnras, 337, 147 
 \bibitem[Matt et al.(2003)]{Matt2003} Matt, G., Guainazzi, M., \& Maiolino, R.\ 2003, \mnras, 342, 422 
\bibitem[McHardy(1985)]{McHardy1985} McHardy, I.\ 1985, \ssr, 40, 559 \bibitem[Mewe et al.(1985)]{Mewe85} Mewe, R., Gronenschild, E.~H.~B.~M., \& van den Oord, G.~H.~J.\ 1985, \aaps, 62, 197 
 \bibitem[Mitsuda et al.(2007)]{Mitsuda07} Mitsuda, K., et al.\ 2007, \pasj, 59, 1 
\bibitem[Mineo et al.(2012)]{Mineo2012} Mineo, S., Gilfanov, M., \& Sunyaev, R.\ 2012, \mnras, 419, 2095 
\bibitem[Murphy \& Yaqoob(2009)]{Mytorus} Murphy, K.~D., \& Yaqoob, T.\ 2009, \mnras, 397, 1549
 \bibitem[Nardini et al.(2015)]{Nardini2015} Nardini, E., Reeves, J.~N., Gofford, J., et al.\ 2015, Science, 347, 860 
 \bibitem[Nardini et al.(2015b)]{Nardini2015b} Nardini, E., Gofford, J., Reeves, J.~N., et al.\ 2015b, \mnras, 453, 2558 
\bibitem[Netzer \& Marziani(2010)]{Netzer-Marziani2010} Netzer, H., \& Marziani, P.\ 2010, \apj, 724, 318 
\bibitem[Netzer(2015)]{Netzer2015} Netzer, H.\ 2015, \araa, 53, 365 
\bibitem[Pereira-Santaella et al.(2011)]{Pereira-Santaella2011} Pereira-Santaella, M., Alonso-Herrero, A., Santos-Lleo, M., et al.\ 2011, \aap, 535, A93 
\bibitem[Piconcelli et al.(2007)]{Piconcelli2007} Piconcelli, E., Bianchi, S., Guainazzi, M., Fiore, F., \& Chiaberge, M.\ 2007, \aap, 466, 855 
  \bibitem[Porquet \& Dubau(2000)]{Porquet2000} Porquet, D., \& Dubau, J.\ 2000, \aaps, 143, 495 
 \bibitem[Ranalli et al.(2003)]{Ranalli2003} Ranalli, P., Comastri, A., \& Setti, G.\ 2003, \aap, 399, 39 
\bibitem[Risaliti et al.(2002)]{Risaliti2002} Risaliti, G., Elvis, M., \& Nicastro, F.\ 2002, \apj, 571, 234 
\bibitem[Risaliti et al.(2005)]{Risaliti05} Risaliti, G., Elvis, M., Fabbiano, G., Baldi, A., \& Zezas, A.\ 2005, \apjl, 623, L93 
\bibitem[Risaliti et al.(2007)]{Risaliti07} Risaliti, G., Elvis,  M., Fabbiano, G., et al.\ 2007, \apjl, 659, L111 
\bibitem[Risaliti et al.(2009)]{Risaliti09} Risaliti, G., et al.\ 2009, \mnras, 393, L1 
\bibitem[Risaliti(2016)]{Risaliti2016} Risaliti, G.\ 2016, Astronomische Nachrichten, 337, 529 
\bibitem[Rivers et al.(2015a)]{Rivers2015a} Rivers, E., Risaliti, G., Walton, D.~J., et al.\ 2015, \apj, 804, 107 \bibitem[Rivers et al.(2015b)]{Rivers2015} Rivers, E., Balokovi{\'c}, M., Ar{\'e}valo, P., et al.\ 2015b, \apj, 815, 55 
\bibitem[Sako et al.(2000)]{Sako00} Sako, M., Kahn, S.~M.,  Paerels, F., \& Liedahl, D.~A.\ 2000, \apjl, 543, L115
\bibitem[Shu et al.(2011)]{Shu2011} Shu, X.~W., Yaqoob, T., \& Wang, J.~X.\ 2011, \apj, 738, 147 
\bibitem[Spergel et al.(2003)]{Spergel2003} Spergel, D.~N., et al.\ 2003, \apjs, 148, 175 
\bibitem[Takahashi et al. (2007)]{Takahashi}Takahashi, T., et  al.\ 2007, \pasj, 59, 35 
\bibitem[Taniguchi \& Ohyama(1998)]{Taniguchi1998} Taniguchi, Y., \& Ohyama, Y.\ 1998, \apjl, 507, L121 
\bibitem[{{Tombesi} {et~al.}(2012){Tombesi}, {Cappi}, {Reeves}, \&  {Braito}}]{Tombesi2012} {Tombesi}, F., {Cappi}, M., {Reeves}, J.~N., \& {Braito}, V. 2012, \mnras, 422,L1
\bibitem[Tombesi et al.(2013)]{Tombesi2013} Tombesi, F., Cappi, M., Reeves, J.~N., et al.\ 2013, \mnras, 430, 1102 
\bibitem[Torricelli-Ciamponi et al.(2014)]{Torricelli-Ciamponi2014} Torricelli-Ciamponi, G., Pietrini, P., Risaliti, G., \& Salvati, M.\ 2014, \mnras, 442, 2116 
\bibitem[Turner et al.(2000)]{Turner2000} Turner, T.~J., Perola, G.~C., Fiore, F., et al.\ 2000, \apj, 531, 245 
\bibitem[Turner \& Miller(2009)]{Turner2009} Turner, T.~J., \& Miller, L.\ 2009, \aapr, 17, 47 
\bibitem[Wang et al.(2009)]{Wang09} Wang, J., Fabbiano, G., Elvis, M., et al.\ 2009, \apj, 694, 718 
\bibitem[Wang et al.(2011)]{Wang2011} Wang, J., Fabbiano, G., Elvis, M., et al.\ 2011, \apj, 736, 62 
\bibitem[Wold et al.(2006)]{Mass_ref} Wold, M., Lacy, M., K{\"a}ufl, H.~U., \& Siebenmorgen, R.\ 2006, \aap, 460, 449 
\bibitem[Xue et al.(1998)]{Xue1998} Xue, S.-J., Otani, C., Mihara, T., Cappi, M., \& Matsuoka, M.\ 1998, \pasj, 50, 519  \bibitem[Yaqoob \& Murphy(2011)]{CSYaqoob} Yaqoob, T., \& Murphy, K.~D.\ 2011, \mnras, 412, 277 
\end{thebibliography}
\end{document}